\documentclass[twocolumn]{aastex631}

\usepackage{amsmath,amssymb,amsthm,amsfonts,latexsym,mathtext}
\usepackage{color}

\def\be{\begin{equation}}
\def\ee{\end{equation}}
\def\ba{\begin{aligned}}
\def\ea{\end{aligned}}

\shorttitle{PeV gamma-ray and neutrino flares from binaries}
\shortauthors{Bykov et al.}
\graphicspath{{./}{fig/}}
\usepackage{ifpdf}
\ifpdf\usepackage{epstopdf}\fi

\def\lsim{\;\raise0.3ex\hbox{$<$\kern-0.75em\raise-1.1ex\hbox{$\sim$}}\;}
\def\gsim{\;\raise0.3ex\hbox{$>$\kern-0.75em\raise-1.1ex\hbox{$\sim$}}\;}
\def\IceC{{\sl IceCube}}

\begin{document}
%\title{PeV gamma-ray emission and neutrino produced by PSR 2032+4127}
\title{PeV photon and neutrino flares from galactic gamma-ray binaries}

%% The \author command is the same as before except it now takes an optional
%% argument which is the 16 digit ORCID. The syntax is:
%% \author[xxxx-xxxx-xxxx-xxxx]{Author Name}
%%
%% This will hyperlink the author name to the author's ORCID page. Note that
%% during compilation, LaTeX will do some limited checking of the format of
%% the ID to make sure it is valid. If the "orcid-ID.png" image file is 
%% present or in the LaTeX pathway, the OrcID icon will appear next to
%% the authors name.
%%
%% Use \affiliation for affiliation information. The old \affil is now aliased
%% to \affiliation. AASTeX v6.31 will automatically index these in the header.
%% When a duplicate is found its index will be the same as its previous entry.

\author{A. M. Bykov}
\affiliation{Ioffe Institute, 26 Politekhnicheskaya, St. Petersburg, 194021, Russia}
%\affiliation{Peter the Great St. Petersburg %Polytechnic University, St. Petersburg, 195251, %Russia}
\author{A. E. Petrov}
\affiliation{Ioffe Institute, 26 Politekhnicheskaya, St. Petersburg, 194021, Russia}
\author{M.E. Kalyashova}
\affiliation{Ioffe Institute, 26 Politekhnicheskaya, St. Petersburg, 194021, Russia}
\affiliation{Peter the Great St. Petersburg Polytechnic University, 29 Politekhnicheskaya, St. Petersburg, 195251, Russia}
\author{S. V. Troitsky}
%\author[0000-0001-6917-6600]{S.~V.~Troitsky}
\affiliation{Institute for Nuclear Research of the Russian Academy of Sciences, 60th October Anniversary Prospect 7a, Moscow, 117312, Russia}

\begin{abstract}
The high-energy radiation from short period binaries containing  a massive star with a compact relativistic companion was detected  from radio to TeV %regime
gamma rays. We show here that PeV regime protons can be efficiently accelerated in the regions of  collision of relativistic outflows of a compact object  with stellar winds in these systems. The accelerated proton spectra in the presented Monte Carlo model have an upturn in the PeV regime and can provide very hard spectra of sub-PeV photons  and neutrinos by photo-meson processes in the stellar radiation field.  Recent report of a possible sub-PeV gamma-ray flare in coincidence with a high-energy neutrino can be understood in the frame of this model. The gamma-ray binaries may contribute substantially to the Galactic component of the detected high-energy neutrino flux.   
\end{abstract}

%\keywords{...}

%\input{intro.tex}
%\input{model.tex}
%\input{discussion.tex}
\section{Introduction}
The cosmic accelerators of petaelectronovolt (PeV) energy particles  revealed themselves through the measured fluxes of the Galactic cosmic rays as well as by the high-energy neutrinos first detected by \IceC   \  \citep[][]{IceCube-2013PRL, IceCube-2013Science}.
%, ANTARES \citep{ANTARES-ICRC2019} and Baikal-GVD %\citep{Baikal-ICRC2021}.  
Ground based $\gamma$-ray telescopes found a population of sources with power-law spectra extending above 10 TeV without a clear signature of the spectral cut-off \citep[][]{2019NatAs...3..561A,GCPeV2016Natur} which can be associated with cosmic pevatrons. Strong attenuation of $\gamma$-ray fluxes above 100 TeV due to photon-photon pair production limits the possibility of the $\gamma$-ray pevatron detection mainly to a population of nearby Galactic sources at distance $D \lsim$ 10 kpc \citep{Nikishov1962,2009herb.book.....D}. The {\sl Large High Altitude Air Shower Observatory}  (LHAASO) collaboration  
reported a significant detection of 12 %ultrahigh-energy 
$\gamma$-ray sources  above 100 TeV and up to 1.4 PeV highlighting the problem of the origin of these PeV accelerators \citep[][]{LHAASONat21}.
The {\sl Carpet-2} experiment team  \citep{Carpet2} recently reported a detection of a 3.1$\sigma$ excess of $\gamma$-ray flux above 300 TeV from the Cygnus region 
associated with a 150-TeV neutrino event detected by \IceC  \      \citep{IceCube-GCN.28927} and
most likely consistent  with a flare of a few months duration. 
The $\gamma$-ray energy flux
%from the source located in the Cygnus region detected by  \citet{Carpet2} 
during the flare is $\gtrsim 10^{-9}$ ${\rm erg~cm^{-2}~s^{-1}}$, an order of magnitude higher than the 95\% CL upper limit on the steady-state flux, $\lesssim 1.2 \cdot 10^{-10}$ ${\rm erg~cm^{-2}~s^{-1}}$, obtained for the same source by {\sl Carpet-2}. The flare flux is well above the fluxes in the TeV regime detected from the population of $\gamma$-ray sources in the region \citep[see e.g.][]{2021PhRvL.127c1102A}. It also highly exceeds the fluxes from known $\gamma$-ray sources including the $\gamma$-ray pulsars, % \citep[e.g.][]{},
supernova remnants \citep[e.g.][]{2021NatAs...5..460T}, superbubbles \citep[e.g.][]{2021NatAs...5..465A},  $\gamma$-ray binaries  \citep[e.g.][]{Aharonian06} as well as from yet unidentified PeV candidate sources \citep[e.g.][]{2021arXiv210609865T,2021arXiv210606405A}. 

Most of high-energy neutrinos are probably extragalactic. Combined {\sl IceCube} and {\sl ANTARES} data \citep{IceCubeANTARES-Gal} limit the Galactic contribution to $\lesssim 15\%$, and first indications to the presence of this component have been found  \citep{IceCube7yrCascades}. Some Galactic contribution helps to explain naturally \citep{2comp-Vissani,NeronovSemikoz2comp} the tension between different measurements of the neutrino spectrum, see e.g.\ \citet{IceCube-HESE2020}. For a review of particular classes of Galactic sources see \citet{Kheirandish-Gal}. Gamma-ray binaries were proposed as high-energy neutrino sources long ago \citep[see e.g.][]{prediction-nu-Waxman2001,prediction-nu-Waxman2002,prediction-nu-Bednarek2005,prediction-nu-Neronov2009,prediction-nu-Neronov2008,prediction-nu-Sahakyan2013}.

The angular resolution of both {\sl Carpet-2} and \IceC \ does not allow one to identify a particular source of the photon flare and of the contemporaneous neutrino in the densely populated Cygnus region. The time variability on a few months timescale and the high observed $\gamma$-ray flux make any associations of the source with extended supernova remnants or superbubbles very unlikely. A number of compact Galactic sources of high-energy radiation are located in the Cygnus region, %just to mention a few well-known -- 
including gamma-ray binaries Cyg X-3 and 
%$\gamma$-ray pulsar 
PSR~J2032+4127.

The maximum energies of %ions with charge number $Z$ 
protons accelerated by outflows with frozen-in magnetic fields of a kinetic/magnetic luminosity ${\cal L_K}$ can be estimated from the equation: 
\begin{equation}\label{Emax}
E_{\rm max} \approx % Z \times 10^{15} \cdot
\frac{{
f\left(\beta_{\rm f}\right)}}{{\rm \Gamma_{\rm f}} \Omega}\left(\frac{
{\cal L_K} }{5\times 10^{34}~\mbox{erg}\,\mbox{s}^{-1}}\right)^{1/2}  \mbox{PeV},
\end{equation}
where the dimensionless velocity of the flow is $\beta_{\rm f} = u_f/c$, $c$ is the speed of light, ${\rm \Gamma}_{\rm f} = 1/\sqrt{1 - \beta_{\rm f}^2 }$ and $\Omega$ is the opening angle of the outflow \citep[see e.g.][and references therein]{2009JCAP...11..009L,B12}. The function $f(\beta_{\rm f}) \propto \beta_{\rm f}^{1/2}$ for $\beta_{\rm f}\ll 1$, while $f(\beta_{\rm f}) \sim 1$ for  ultra-relativistic flow with ${\rm \Gamma}_{\rm f} \gg 1$. It follows from the equation that for a given ${\cal L_K}$, the higher values of $E_{\rm max}$ can be achieved for the mildly relativistic flows with $\beta_{\rm f} {\rm \Gamma}_{\rm f} \sim 1$ with relatively narrow opening angle $\Omega < 1$. Then ${\cal L_K} > 3 \times  10^{35}~\mbox{erg}\, \mbox{s}^{-1}$ and $\Omega < 1/3$ are needed to reach the energy of the accelerated proton $\sim$ 10 PeV. Eq. (\ref{Emax}) does not account for radiative losses, which can reduce $E_{\rm max}$ substantially. 

%The PeV-band $\gamma$-rays cannot propagate for distance much %longer than about $\sim 10$ kpc due to the pair production on %the Cosmic Microwave Background \citep[see %e.g.][]{Nikishov1962,2009herb.book.....D}, 
%%unless some unconventional physics is involved \citep[e.g.\ %axion-like particles,][]{ST-ALPgammaRev}. Therefore 
%so the source detected by {\sl Carpet-2} in the Cygnus %direction is very likely Galactic. 

The observed flare flux transforms to the $\gamma$-ray luminosity above 300 TeV of  $L_{\gamma} \sim 4\times 10^{35} (D/1.5 \,\mbox{kpc})^2$ erg\,s$^{-1}$. Thus, if the source is indeed located in the Cygnus star-forming region then, according to Eq. (\ref{Emax}), ${\cal L_K}$ required to produce PeV regime particles radiating 300-TeV photons is comparable to $L_{\gamma}$. 

One can conclude that a very hard spectrum of accelerated particles and a fast cooling of PeV particles are needed in order the required kinetic/magnetic luminosity to be consistent with that available in compact relativistic sources of stellar masses. We present below a model of the compact $\gamma$-ray sources that can convert a substantial fraction of their kinetic luminosity (provided by the magnetic braking of pulsar, magnetic field reconnection in magnetars, or accretion onto black hole) to the PeV regime $\gamma$-rays and neutrinos by photo-meson production mechanism in proton-photon collisions. Gamma-ray binary sources (LS 5039,
PSR B1259-63, LSI +61° 303, PSR~J2032+4127, and others, see e.g.\ \citet{2013A&ARv..21...64D}) can be considered as possible candidates. Well-known powerful microquasar Cyg X-3 demonstrated giant $\gamma$-ray flares \citep[][]{CorbelCygX3flare12}. In all of these sources, which have been subjects of extensive modeling for a long time \citep[see e.g.,][]{1997ApJ...477..439T,2020MNRAS.497..648C}, the compact object has a massive early type star companion.   

The model of a PeV source we discuss here suggests the interaction of a fast outflow from a compact relativistic object with the stellar wind (SW) of a bright massive star. The colliding magnetized flows provide a plausible way to a fast Fermi-type acceleration to the PeV regime of TeV-energy particles pre-accelerated in the vicinity of the compact object. The acceleration mechanism at the colliding wind flows (CWFs) may form very hard spectrum of particles in the TeV-PeV energy range.
In addition to $\gamma$-ray radiation produced in proton-proton collisions, the accelerated PeV protons interact efficiently with the optical photons of the luminous massive star by photo-meson mechanism providing fast cooling in PeV $\gamma$-rays and neutrinos.
In the next section we will discuss the PeV $\gamma$-ray and neutrino production in the generic source  PSR~J2032+4127, though a similar model could be applied to other binary sources including either pulsar winds (PWs) or jets of the accreting black holes.

\section{Model}
PSR~J2032+4127 is located at distance $\sim$ 1.4 kpc and orbits around a massive Be star MT91 213 (B0Vp) 
%\citep{CygOB2_stars91,2015MNRAS.451..581L}
with a long period $\sim$ 50 yr \citep{2015MNRAS.451..581L,2017MNRAS.464.1211H}. %is a young radio-loud pulsar emitting in $\gamma$-rays \citep{2009Sci...325..848A,2009ApJ...705....1C}. 
Its spin-down power $\dot{E}$ may reach 3$\times$10$^{35}$\,erg\,s$^{-1}$ \citep[][]{2009ApJ...705....1C}.
%in case of a stiff equation of state of the %neutron star matter allowing the moment of %inertia $\sim 2 \times 10^{45}$ g\,cm$^2$. 
This is close to the $\gamma$-ray luminosity $\sim $ 2.7$\times$10$^{35}$\,erg\,s$^{-1}$ at the lower limit of {\sl Carpet-2} measurement uncertainty band. This suggests that most of the CWFs acceleration source power should be converted into PeV range protons energy. The total available power of the CWF is a sum of the pulsar spin-down power and the fraction of the SW power released in the CWFs.    
Fermi I type acceleration in the CWFs can produce particle energy distribution $f\left(E\right) \propto E^{-s}$ with $s \sim$ 1, where energy is mainly accumulated by the most energetic particles \citep[see][]{BSPWN_2017}. Efficient acceleration at CWFs may be caused by passage of the pulsar through the equatorial region of the Be star SW. In November 2020 the pulsar was at $\sim 20$ AU from MT91 213 that is about $\sim$ 400 stellar radii allowing interaction of the PW with the Be star equatorial disk \citep{Be_disk17}.

The model of PeV flares of PSR~J2032+4127 we discuss here uses only the hadronic emission processes.
Severe synchrotron radiation losses do not allow acceleration of PW leptons to PeV energies at considered orbital phase. %at periastron stage of considered binary.
%PeV electron cooling timescale in the magnetic field $\sim$ 1 G is much shorter than the fastest acceleration timescale $\sim r_g / c$, where $r_g$ is the particle gyroradius. 
The toroidal SW magnetic field component scaling $\propto r^{-1}$ gives $\sim$ 1 G at distance $r \sim $ 20 AU, if the surface dipolar stellar field is $\sim$ 1 kG. For a fast rotating star with mass $M_{\ast} \sim 15 M_{\odot}$ and radius $R_{\ast} \sim 10 R_{\odot}$, even higher field $\sim$ 2 kG may be expected \citep[see, e.g.,][]{2019MNRAS.490..274S}.     
%Indeed, given the surface dipolar field $\sim$ 0.1-1 kG, one expects the radial component of SW magnetic field scaling $\propto r^{-2}$ to be $\sim$ 0.2-2 G at the periastron orbital separation distance $r =$ 1 AU. As we discuss further, actual fields at the particle acceleration site may be even stronger. In the 0.2 G field an electron with 0.3 PeV energy loses half of it due to synchrotron losses in $t_{syn} \approx 5 \times 10^8 / B^2 \gamma \sim 20$ seconds, whereas acceleration requires at least $\sim r_g / c  \sim 170$ seconds, where $\gamma$, $r_g$ and $c$ are the particle Lorentz-factor, particle gyroradius and the speed of light, respectively. Thus, we should indeed consider the hadronic nature of the PeV emission.

%Stellar wind protons transported to the winds interaction region first may be accelerated in the Fermi I type process at the front of the bow shock. Then they may be injected in the Fermi I type acceleration in the colliding shock flows. 
%Let us assume that the particles accelerated to highest energies have gyroradii of order of the bow shock standoff distance $R_{sh}$ which is of order of the pulsar and star orbital separation at the periastron $\sim$ 1 AU. Then the timescale of the entire acceleration process may be estimated as $t_{acc} \sim R_{sh} / u$, where $u$ is the velocity of upstream stellar wind flow in the pulsar rest frame. For $u \sim 10^7$ km\,s$^{-1}$ and $R_{sh} = 10^{13}$ cm one obtains $t_{acc} \sim 10^6$ s -- much smaller than the observed flare duration $\sim 8 \times 10^6$ s.   

\begin{figure}%[h!]
\center{\includegraphics[width=1.0\linewidth]{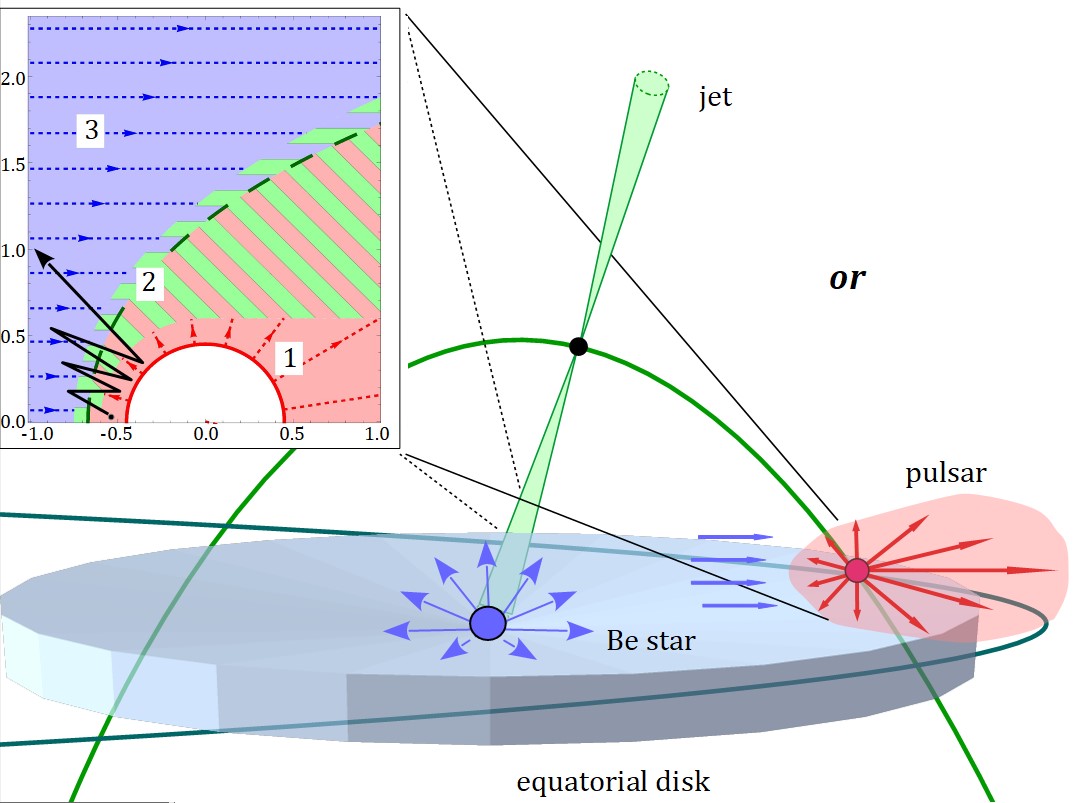}}
%\center{\includegraphics[width=1.0\linewidth]{fi%g/regions_PSR_2032_sketch_nums.pdf}}
\caption{\label{fig:sketch_geometry}Sketch of the interaction of mildly relativistic flows produced by a compact object (pulsar or black hole) with the equatorial disk of Be star. Inset: spatial structure of the Monte Carlo model diffusion zones. Pink zone (1) -- shocked pulsar wind, blue (3) -- stellar wind, cross-hatched with green (2) -- zone around the contact discontinuity (approximate position -- dashed green line).
White zone is the cold pulsar wind.
Red solid line shows the termination shock of the pulsar wind, dashed lines with arrow are the directions of flows. Lengths are normalized to $2.5 \times 10^{14}$ cm. The black imposed polygonal line illustrates a trajectory of a particle accelerated in the colliding flows.}
\end{figure}

We simulate acceleration at the CWFs using the kinetic Monte Carlo model described in Section 4 of \cite{BSPWN_2017} adapted for this problem. This model allows simulating the diffusive particle propagation 
in the region where the PW collides with the ambient matter flow, launching the PW termination shock and possibly the bow shock.
The model plausibly catches the spatial structure of CWFs system represented by spatial zones with reasonable parameters of diffusion and magnetic fields and includes a simple model of flows velocities distribution (see Fig. \ref{fig:sketch_geometry}).

\label{sketch}

Protons are injected in the CWFs region with a wide soft power-law spectrum $f_{\rm inj}\left(E\right) \propto E^{-s}$ with $s \sim 2.2$ \citep[e.g.][]{B12,2015SSRv..191..519S,2020arXiv200104442A}. It is produced by Fermi I type acceleration at the  termination shock formed in the collision of the pair-dominated pulsar wind with magnetized plasma of stellar wind. Both pairs and protons have to be accelerated in this system. In the simulation we generate a population of particles with the energy distribution function $f_{\rm inj}\left(E\right)$ and inject them in the CWFs at the contact discontinuity (see Fig.~\ref{fig:sketch_geometry}).

%Some small fraction $\zeta$ of the SW proton flux is injected in the Fermi acceleration at the BS which produces a wide soft power-law spectrum $f_{\rm inj}\left(E\right) \propto E^{-s}$ with $s \sim 2$ in the energy range $E_{\rm min} < E < E_{\rm max}$. Low energy bound lies at $E \sim$ 1 GeV, while the maximum energy should be defined by matching particle gyroradius and accelerator spatial scale ($E_{\rm max}^{\rm acc}$) or acceleration and radiative energy losses timescales ($E_{\rm max}^{\rm rad}$) so that $E_{\rm max} = \min\left(E_{\rm max}^{\rm acc}, E_{\rm max}^{\rm rad}\right)$.

%Particles accelerated at the BS front are injected in the Fermi I type acceleration in the CWFs simulated by our Monte Carlo code. In the simulation run we generate a population of particles with the energy distribution function $f_{\rm inj}\left(E\right)$ and inject them in the CWFs at the inner bound of the Region 3 in Fig.~\ref{fig:sketch_geometry}.

Particles then propagate through the CWFs according to the adopted diffusion model. The diffusion coefficients %in the vicinity of the shocks as well as at the contact discontinuity
are chosen in the Bohm limit, i.e., defined by the particle energy and local magnetic field. 
Each particle propagates to a distance defined by its mean free path (mfp) given by the diffusion coefficient at its location, then it is scattered isotropically in the local plasma rest frame, then again propagates to a mfp distance --- and so on, until it leaves the simulation. All the generated particles are propagated through the simulation area one by one. Multiple subsequent scatterings and crossings of the contact discontinuity in the zone of almost head-on winds collision allows particle to gain energy. Particles can be accelerated until either their mfp exceeds the size of accelerator allowing them to escape through the border with free escape boundary condition, or they lose too much energy due to adiabatic and radiative losses and therefore join the low-energy particle pool. 

Relativistic protons radiate mainly due to interactions with stellar photons (photo-meson process, also dubbed as `$p\gamma$') and SW protons ($pp$ process). The radiative losses (including synchrotron and inverse Compton) are accounted for at each scattering by subtraction of the energy radiated by a particle in all the radiation processes during the preceding free propagation flight. We calculate the radiation loss rates using the approaches of \citet{2006PhRvD..74c4018K,2014PhRvD..90l3014K} for the $pp$ process and \citet{2008PhRvD..78c4013K} for the $p\gamma$ process.

%Detection of momenta of all particles leaving the simulation box with the Monte Carlo techniques allows to get the spectrum of particles accelerated by the CWFs. The simulation allows to get the accelerated particles energy distribution, averaged over the simulation volume. 
Detection of momenta of all particles leaving the simulation box with the Monte Carlo techniques allows us to get the spectral energy distribution of particles accelerated by the CWFs, averaged over the simulation volume. 
Then, the emission produced in $p\gamma$ and $pp$ processes 
is calculated using parametrizations of \citet{2006PhRvD..74c4018K,2008PhRvD..78c4013K,2014PhRvD..90l3014K}. 
We take into account the finite size of the simulation box, scaling the photon field $\propto r^{-2}$ with distance from the star $r$. To do so, we divide the box into a number of spatial bins, calculate the emission flux produced by each bin separately and finally summarize the results.

For a thermal distribution of stellar photons with $T \approx$ 30000~K, the $p\gamma$ process %thought to produce PeV gamma-ray flux observed by Carpet
works above a threshold proton Lorentz-factor $\Gamma_p \gsim 10^7$ \citep[e.g.][]{2009herb.book.....D}.
Thus, to enable the $p\gamma$ process for the peak of CWF-produced hard energy distribution, the peak should lie at $\gsim$ 10 PeV.
%Thus, to let particles from the peak of CWF-produced hard energy distribution radiate in $p$-$\gamma$-process most of the available spin-down power, colliding flows should confine protons with Lorentz-factors upto $\sim 10^7$ during the acceleration.
%To reproduce the observed PeV energy flux without exceeding TeV energy flux a hard spectrum with $s \sim 1$ is required. Such a spectrum may be produced in Fermi acceleration in CWFs region, where particles with Lorentz-factor up to 10$^7$ should be confined during the acceleration process.

%Orbital distance between the massive star and the pulsar during its periastron passage is about 1 AU. Thus, the bow shock apex is limited by $R_{sh} \sim 10^{13}$ cm. 

The structure of the outer disks of Be stars was studied by \citet{Be_disk17}. They estimated the outer disk extensions to be up to $\sim$ 400 equatorial radii of the star and found that the disk height $H$ scales as $H \propto r^{3/2}$. This means that the disk could extend to the distance of about 20 AU with the width well above 1 AU.

Maximum standoff distance $R_{\rm sd} = \sqrt{\dot{E}/4\pi\rho u^2 c} \sim $ 10 AU defining the CWFs acceleration region size is limited by the orbital separation distance. Here $\rho$ and $u$ are the SW mass density and velocity at the winds interaction region.
Confinement of a particle with a gyroradius $r_g$ in the CWFs, $r_g \lsim R_{\rm sd}$,
%A particle confinement (particle gyroradius $r_g \lsim R_{\rm sd}$) 
then requires the magnetic field $\gsim$ 1 G. For the surface dipolar field of the star $\sim$ 1 kG the toroidal SW field allows $\sim$ 2 G at $r \sim 2 \times 10^{14}$ cm.
%Such a field indeed may be provided by a massive star with mass $M_{\ast} \sim 15 M_{\odot}$ and radius $R_{\ast} \sim 10 R_{\odot}$. According to \cite{2019MNRAS.490..274S}, such a star may have the surface dipolar field $\sim$ 2 kG. Toroidal component of SW magnetic field %scaling $\sim r^{-1}$ with distance from the star $r$ 
%then allows $\sim$ 4 G at distance $r \sim 2 \times 10^{14}$ cm. 
%To figure out the actual BS region field one has to resolve uncertainties related to the actual surface magnetic field of MT91 213 and geometry of the winds interaction, e.g., the obliquity of the SW field with respect to the shock that governs field jump. Keeping in mind these ambiguities, we assumed the BS region field to be $\sim$ 50 G. 
%%Assuming that such a star produces stellar wind with toroidal field $\sim$ 400 G at launch and falling $\propto r^{-1}$, one expects $\sim$ 50 G at bow shock apex.

%Magnetic field in the shocked PW $B_{\rm pwn}$ may be estimated using results of \citet{1984ApJ...283..694K}. Assuming wind magnetization $\sim$ 0.01 and termination shock radius $\sim 0.5 R_{\rm sh}$ one may obtain $B_{\rm pwn} \sim$ 200 kG in the plasma rest frame.
%Following \citet{2017ApJ...836..241T}, we estimated the relative velocity of SW and pulsar as $u \sim 2 \times 10^7$ km\,s$^{-1}$.
At $r \gg R_{\ast}$ radial component of SW velocity is about its terminal velocity $v_{\infty}\sim \sqrt{2 G M_{\ast} / R_{\ast}} \approx 800 \,\mbox{km}\,\mbox{s}^{-1}$, where $G$ is the gravitational constant. This is much faster than the orbital velocity $v_K \sim 30 \,\mbox{km}\,\mbox{s}^{-1}$ and the wind toroidal velocity $\sim$ few $\mbox{km}\,\mbox{s}^{-1}$. Thus, we assume the SW velocity $u \approx v_{\infty}$ and SW protons number density $\sim$ 3000 cm$^{-3}$ corresponding to the SW disk density at stellar surface $\sim 10^{-12} \,\mbox{g}\,\mbox{cm}^{-3}$ \citep{Be_disk17}. 
%The number density of the SW protons estimated from the required $R_{\rm sh} \sim 10^{13}$ cm is about $10^7$ cm$^{-3}$, assuming $\dot{E} = 3.4 \times 10^{35}$ erg\,s$^{-1}$. %(for a stiff equation of state and moment of inertia of the neutron star $\sim 2 \times 10^{45}$ g\,cm$^2$).

The timescale of the acceleration in the CWFs is much shorter than flare duration $\sim 10^7$ s. According to Eq. (21) from \citet{BSPWN_2017} it takes $\sim 10^6$ s to produce a hard energy distribution peaking at tens of PeV. The CWFs acceleration may provide the high enough value of $\gamma$-ray flux during the pulsar passage through the disk that also may take $H / u \sim 10^7$ s.
%The particle acceleration due to Fermi I type processes in the CWFs is much faster than the duration of flare. Particles accelerated even up to 100 PeV would have gyroradii $r_g \sim 6 \times 10^{12}$ cm in $B_{\rm bow} \sim 50$ G.
%Acceleration timescale of the entire system may be estimated as $t_{acc} \sim r_g / u \approx 3 \times 10^5$ s, much shorter than $\sim 90$~d$ \approx 8 \times 10^6$ s, allowing to consider production of hard spectrum as a stationary process. This allows us to treat the BS and CWFs as a stable accelerating engine working during the flare and giving an average power of about full pulsar spin-down.

%\begin{figure}[h!]
\begin{figure}
%\center{\includegraphics[width=1.0\linewidth]{CRs.eps}}
\center{\includegraphics[width=1.0\linewidth]{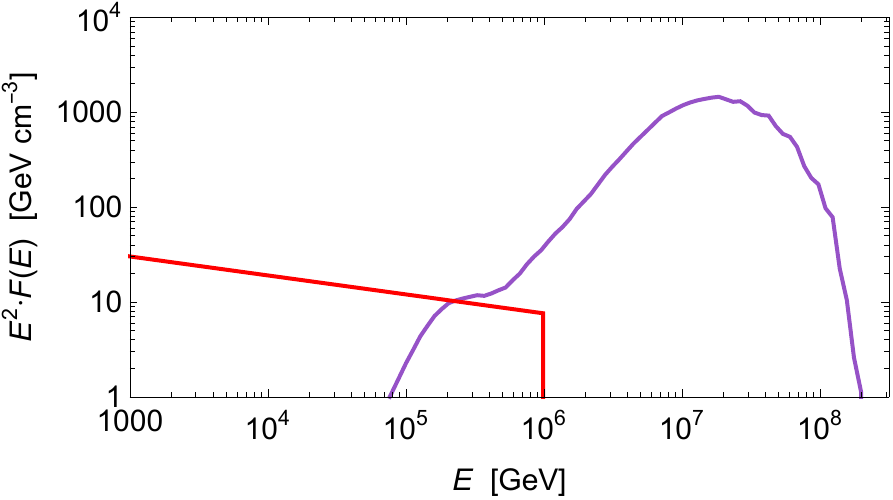}}
\caption{\label{fig:particles}The energy distribution function of particles injected into the colliding wind flows (red; extends down to 1 GeV) and the result of Monte Carlo simulation of spectrum of particles accelerated in the colliding wind flows in the collision zone of pulsar and stellar winds (violet).}
\end{figure}\label{protons}

%\begin{figure}[h!]
%\center{\includegraphics[width=1.0\linewidth]{flux.eps}}
%\caption{\label{fig:emission}Simulated p-p %(dotted red) and p-$\gamma$ (dashed red) %emission flux produced by the simulated %distribution of accelerated protons. Solid red %curve shows the sum of p-p and p-$\gamma$ %components. Green line and shaded area %represents the flux of gamma-ray emission %detected by Carpet-2, while blue and pink shaded %areas in the left bottom corner -- the %uncertainties areas for {\sl VERITAS} and {\sl %MAGIC}.}
%\end{figure}\label{gamma}

%\begin{figure}[h!]
\begin{figure}
%\center{\includegraphics[width=1.0\linewidth]{flux_band.eps}}
\center{\includegraphics[width=1.0\linewidth]{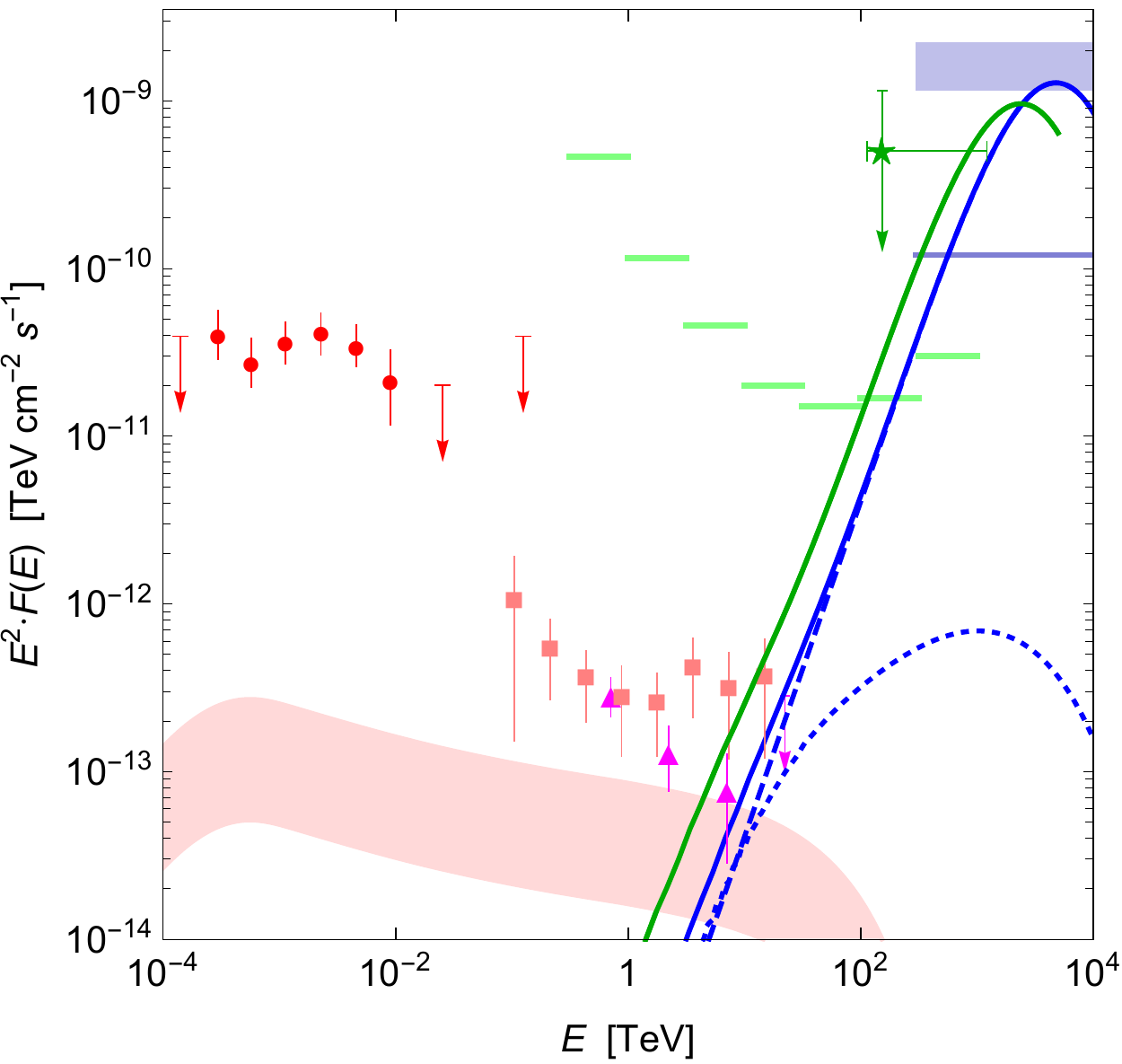}}
\caption{\label{fig:emission}High-energy emissions from the source in the Cygnus region. Blue -- photons: total (full curve) simulated flare flux, contributions from p-p (dotted), p-$\gamma$ (dashed) interactions of protons accelerated in colliding winds;  red shaded area: photons from p-p interaction of injection spectrum protons (within the source parameters uncertainties); violet shaded area: {\sl Carpet-2} flare flux;
magenta and pink data points: {\sl VERITAS} and {\sl MAGIC} steady-state fluxes \citep{VERITAS_MAGIC}, 
%red and gold -- their steady-state fluxes; 
red points in the GeV range -- {\sl Fermi LAT} steady-state data \citep{2020MNRAS.495..365C}; violet horizontal line: {\sl Carpet-2} 95\% CL upper limit on the steady-state flux. Green -- neutrinos: estimated total flare flux (full curve); 
 horizontal lines: 90\% CL {\sl IceCube} \citep{IceCube-binaries} upper limit on the steady-state neutrino flux of Cyg~X-3, expected to be similar to those of PSR~J2032+4127; the star: an order-of-magnitude estimate \citep{Carpet2} of the flare flux from the detection of  one 150-TeV neutrino \citep{IceCube-GCN.28927}.
}
\end{figure}\label{gamma}

\section{Results and discussion}

Figure \ref{fig:particles} presents the simulated spectrum of accelerated protons. The red curve shows the injection spectrum, the purple curve --- particle spectrum after the CWFs acceleration. The latter is much harder and peaks at a few tens of PeV, where most of particle energy is accumulated.

In Figure \ref{fig:emission} we show the simulated emission spectra  produced by the modeled particle distribution. The $pp$ gamma-ray flux (dotted curve) is well below the {\sl Carpet-2} measured flux, whereas $p\gamma$ flux (dashed curve) produced by particles accelerated up to tens PeV allows us to explain the results of {\sl Carpet-2} flare measurements. At TeV energies the total flux produced in both hadronic processes does not exceed even the steady-state fluxes observed by {\sl VERITAS} and {\sl MAGIC} \citep{VERITAS_MAGIC}.
%{\bf 
The Fermi LAT 1-10 GeV lightcurve of PSR J2032+4127 is stable along the whole orbit with the flux $\sim$ 10$^{-10}$ erg cm$^{-2}$ s$^{-1}$ \citep[][]{2020MNRAS.495..365C}. It is likely produced by the pulsar magnetospheric emission and the flux is well above that from hadronic interactions described by our model, see Fig.~\ref{fig:emission}. The GeV-TeV spectra are successfully modelled within the leptonic scenario \citep[see e.g.][]{2020MNRAS.495..365C,2017ApJ...836..241T}.
%}
As it was mentioned above, severe radiation losses exclude the leptonic origin of PeV gamma-ray emission.
Interestingly, the indications of leptons spectral break from $s \sim 2$ to a harder $s \sim 1$, similar to shown in Fig.~\ref{fig:particles} for protons spectrum, were obtained in the hard X-ray spectrum of LS 5039 detected with  {\sl IBIS INTEGRAL} \citep{falanga21}.
%Interestingly, similarly to our modeled proton spectrum (Fig.~\ref{fig:particles}), the indications of spectral breaks of leptons from index $s \sim 2$ to a harder $s \sim 1$ were obtained in the hard X-ray spectrum of LS 5039 detected with {\sl IBIS INTEGRAL} \citep{falanga21}.

The $\gamma$-ray binaries may represent a possible
population of cosmic ray sources with hard spectra. This can be important, e.g., to resolve the issues with energetics of high-energy radiation %discussed by
\citep[see][]{2019PhRvD..99f3012M}.
%A presence of additional cosmic-ray sources with hard spectra (Galactic transients or extragalactic) was discussed by  \citet{2019PhRvD..99f3012M} to resolve the issues with energetics of high-energy radiation.
%The gamma-ray binaries may therefore represent a possible population.
%The gamma-ray binaries discussed here as possible PeV flares sources may also represent a possible population.

The effects of the absorption of GeV-TeV $\gamma$-rays in close $\gamma$-ray binaries containing luminous massive stars were discussed in the case of LS 5039, LSI +61° 303 and Cyg X-3 \citep[see, e.g.,][]{2006MNRAS.368..579B,2011A&A...529A.120C}. However, in the source considered above, the attenuation of PeV photons of interest due to e$^{\pm}$ pair production on the intensive stellar radiation  is not essential.

The predicted neutrino flux is compared in Figure \ref{fig:emission} with typical estimates of the flare and steady-state neutrino emission from {\sl IceCube} data. Like in $\gamma$-rays, the flare neutrino flux is order of magnitude higher than the steady one. The overall contribution of this and similar sources to the {\sl IceCube} astrophysical neutrino flux is limited by short duty cycles and by a small number of sources. We estimate that the time-averaged flux from all gamma-ray binaries can reach at most $\sim (10-13)\%$ of the full-sky astrophysical neutrino flux, in agreement with constraints on the diffuse Galactic component \citep{IceCubeANTARES-Gal}; studies of point-source contributions  \citep[e.g.,][]{IceCube-binaries} are less constraining yet.

In the PeV source model we considered a particular geometry of the mildly relativistic colliding flows of PW -- massive Be star wind shown in Fig.~\ref{fig:sketch_geometry}. 
The same mechanism can accelerate the protons rapidly up to PeV energies in other $\gamma$-ray binaries, e.g. in case of the black hole jet colliding with the massive star wind also illustrated in Fig.~\ref{fig:sketch_geometry}.
The realistic models of interaction of the fast outflow from a compact object with the wind of a massive star are still under debate.
The models of LS 5039 with anisotropic PW  \citep{bosch-ramon21} and microquasars like Cyg X-1 and  Cyg X-3 with relativistic jets from the accreting stellar mass
black holes \citep[see, e.g.,][and the references therein]{Romero17} represent a few of these.
%The anisotropic PW models of LS 5039 \citep[e.g.][]{bosch-ramon21} and relativistic jets of the microquasars due to accretion onto the stellar mass black holes like Cyg X-1 and Cyg X-3 \citep[see, e.g.,][and the references therein]{Romero17} represent a few of these.
Apart from the microquasars and pulsar wind nebulae \citep[e.g.,][]{2012SSRv..173..341A,2020arXiv200104442A} the powerful outflows of the high enough kinetic luminosities (consistent with that given by  Eq.\ref{Emax}) in supernova remnants \citep{2021Univ....7..324C}, compact stellar clusters \citep{2015MNRAS.453..113B} and superbubbles  can accelerate protons above PeV and produce (sub-)PeV $\gamma$-rays and neutrino. 
The distinctive feature of the model of the $\gamma$-ray binary PSR~J2032+4127 --- MT91 213 
discussed above is the combination of both the fast production of hard energy spectra of protons peaking in the PeV regime and the fast efficient photo-meson cooling in the vicinity of the B0Vp star which provides the highest PeV regime luminosity.
\\
\\
\\
\section{Conclusions}
The model we discuss in the paper allows converting a sizeable fraction of the kinetic power of the outflows of the compact objects in close binary systems to PeV regime radiation due to the high efficiency of Fermi mechanism in colliding flows. % \citep[see e.g.][]{BSPWN_2017}.
Recent modeling by \citet{Pittard21} of TeV regime particle acceleration in colliding wind binary with wind velocities $\sim \mbox{few}\times10^3$ $ \mbox{km}\,\mbox{s}^{-1}$ and $\sim$ mG magnetic fields in the acceleration region demonstrated that $\sim$ 30\% of the wind power was transferred to non-thermal particles. In our model with the short binary separation allowing for $\gsim$ G magnitude magnetic fields and mildly relativistic flows produced by shocked relativistic outflow of a compact object one can reach a similarly high efficiency but for proton acceleration to PeV energy regime.
The rapid photo-meson cooling of the PeV protons 
%allows converting a significant fraction of the kinetic power 
converts a significant part of the available kinetic power
to (sub-)PeV photons and neutrinos.

Our model generally predicts a transient character of 
the very bright PeV regime radiation. 
%This is because of two reasons. 
The main reason of that is the presence of the variations of the magnetic field, particle and seed photon densities along the orbit of the compact object due to the disks (for a Be companion star) or anisotropic stellar winds. This provides variability of the very-high-energy radiation along the compact object orbit. The timescale to cross the Be star disk could be a few months for the orbital parameters of PSR~J2032+4127 which may explain the estimated duration of the $\gamma$-ray  flare detected by {\sl Carpet-2}.   

%The threshold character of the photo-meson radiation mechanism is also important.
An important factor is also the threshold character of the photo-meson radiation mechanism. %which provides a fast cooling of protons responsible for the very bright emission of the PeV regime photons and neutrinos.
Indeed, the energy of a seed photon in the rest frame of the accelerated proton must exceed $\sim$ 200 MeV to start the  %photo-meson radiation 
mechanism, which requires protons of energies above 10 PeV to interact with the intense optical radiation of the massive star.  

A transient activation of the efficient and fast Fermi I type acceleration up to tens of PeV in the colliding flows during passage of the compact companion through the equatorial disk and rapid photo-meson cooling allows the interpretation
%provides a (sub-)PeV photon and neutrino flare. 
of the very bright PeV photon flare detected by {\sl Carpet-2}. Moreover, our mechanism unavoidably provides high neutrino flux, which could be detected by {\sl IceCube} and other observatories.
PeV neutrinos produced by the gamma-ray binaries in starforming galaxies may contribute to the observed population of the very high energy neutrinos.     

%\purple{The protons acceleration to PeV regime is expected in other Galactic objects such as the compact stellar clusters (CSC), superbubbles and young pulsar wind nebulae \citep[see, e.g.,][and the references therein]{2021Univ....7..324C}. The photo-meson production of the PeV gamma-rays may operate in the CSCs, however, one cannot expect a flaring character of their emission. Other sources does not provide either a hard particle spectra or an intense seed optical photons field.}

%A presence of additional cosmic-ray sources with hard spectra (Galactic transients or extragalactic) was discussed by  \citet{2019PhRvD..99f3012M} to resolve the issues with energetics of high-energy radiation. The flaring gamma-ray binaries may represent a possible population.

\section*{Acknowledgments} 
       We thank the referee for useful comments.
       A.M.B. and A.E.P. were supported by the RSF grant 21-72-20020. M.E.K. was supported by RFBR grant 20-32-90156. S.V.T. was supported by the Ministry of science and higher education of Russian Federation under the contract 075-15-2020-778. Some of the modeling was performed at the Joint Supercomputer Center JSCC RAS and at the ``Tornado'' subsystem of the St.~Petersburg Polytechnic University supercomputing center.

%\appendix
%\input{diffusion_model.tex}


\bibliography{PSR_2032_bib}{}

\begin{thebibliography}{}
\expandafter\ifx\csname natexlab\endcsname\relax\def\natexlab#1{#1}\fi
\providecommand{\url}[1]{\href{#1}{#1}}
\providecommand{\dodoi}[1]{doi:~\href{http://doi.org/#1}{\nolinkurl{#1}}}
\providecommand{\doeprint}[1]{\href{http://ascl.net/#1}{\nolinkurl{http://ascl.net/#1}}}
\providecommand{\doarXiv}[1]{\href{https://arxiv.org/abs/#1}{\nolinkurl{https://arxiv.org/abs/#1}}}

\bibitem[{{Aartsen} {et~al.}(2013){Aartsen}, {Abbasi}, {Abdou}, {Ackermann},
  {Adams}, {Aguilar}, {Ahlers}, {Altmann}, {Auffenberg}, {Bai}, {Baker},
  {Barwick}, {Baum}, {Bay}, {Beatty}, {Bechet}, {Becker Tjus}, {Becker},
  {Bell}, {Benabderrahmane}, {BenZvi}, {Berdermann}, {Berghaus}, {Berley},
  {Bernardini}, {Bernhard}, {Bertrand}, {Besson}, {Binder}, {Bindig}, {Bissok},
  {Blaufuss}, {Blumenthal}, {Boersma}, {Bohaichuk}, {Bohm}, {Bose},
  {B{\"o}ser}, {Botner}, {Brayeur}, {Bretz}, {Brown}, {Bruijn}, {Brunner},
  {Carson}, {Casey}, {Casier}, {Chirkin}, {Christov}, {Christy}, {Clark},
  {Clevermann}, {Coenders}, {Cohen}, {Cowen}, {Cruz Silva}, {Danninger},
  {Daughhetee}, {Davis}, {De Clercq}, {De Ridder}, {Desiati}, {de With},
  {DeYoung}, {D{\'\i}az-V{\'e}lez}, {Dunkman}, {Eagan}, {Eberhardt}, {Eisch},
  {Ellsworth}, {Euler}, {Evenson}, {Fadiran}, {Fazely}, {Fedynitch},
  {Feintzeig}, {Feusels}, {Filimonov}, {Finley}, {Fischer-Wasels}, {Flis},
  {Franckowiak}, {Franke}, {Frantzen}, {Fuchs}, {Gaisser}, {Gallagher},
  {Gerhardt}, {Gladstone}, {Gl{\"u}senkamp}, {Goldschmidt}, {Golup},
  {Gonzalez}, {Goodman}, {G{\'o}ra}, {Grant}, {Gro{\ss}}, {Gurtner}, {Ha}, {Haj
  Ismail}, {Hallen}, {Hallgren}, {Halzen}, {Hanson}, {Heereman}, {Heinen},
  {Helbing}, {Hellauer}, {Hickford}, {Hill}, {Hoffman}, {Hoffmann}, {Homeier},
  {Hoshina}, {Huelsnitz}, {Hulth}, {Hultqvist}, {Hussain}, {Ishihara},
  {Jacobi}, {Jacobsen}, {Jagielski}, {Japaridze}, {Jero}, {Jlelati},
  {Kaminsky}, {Kappes}, {Karg}, {Karle}, {Kelley}, {Kiryluk}, {Kislat},
  {Kl{\"a}s}, {Klein}, {K{\"o}hne}, {Kohnen}, {Kolanoski}, {K{\"o}pke},
  {Kopper}, {Kopper}, {Koskinen}, {Kowalski}, {Krasberg}, {Krings}, {Kroll},
  {Kunnen}, {Kurahashi}, {Kuwabara}, {Labare}, {Landsman}, {Larson},
  {Lesiak-Bzdak}, {Leuermann}, {Leute}, {L{\"u}nemann}, {Madsen}, {Maruyama},
  {Mase}, {Matis}, {McNally}, {Meagher}, {Merck}, {M{\'e}sz{\'a}ros}, {Meures},
  {Miarecki}, {Middell}, {Milke}, {Miller}, {Mohrmann}, {Montaruli}, {Morse},
  {Nahnhauer}, {Naumann}, {Niederhausen}, {Nowicki}, {Nygren}, {Obertacke},
  {Odrowski}, {Olivas}, {Olivo}, {O'Murchadha}, {Paul}, {Pepper}, {P{\'e}rez de
  los Heros}, {Pfendner}, {Pieloth}, {Pinat}, {Pirk}, {Posselt}, {Price},
  {Przybylski}, {R{\"a}del}, {Rameez}, {Rawlins}, {Redl}, {Reimann}, {Resconi},
  {Rhode}, {Ribordy}, {Richman}, {Riedel}, {Rodrigues}, {Rott}, {Ruhe},
  {Ruzybayev}, {Ryckbosch}, {Saba}, {Salameh}, {Sander}, {Santander}, {Sarkar},
  {Schatto}, {Scheel}, {Scheriau}, {Schmidt}, {Schmitz}, {Schoenen},
  {Sch{\"o}neberg}, {Sch{\"o}nwald}, {Schukraft}, {Schulte}, {Schulz},
  {Seckel}, {Sestayo}, {Seunarine}, {Sheremata}, {Smith}, {Soiron}, {Soldin},
  {Spiczak}, {Spiering}, {Stamatikos}, {Stanev}, {Stasik}, {Stezelberger},
  {Stokstad}, {St{\"o}{\ss}l}, {Strahler}, {Str{\"o}m}, {Sullivan}, {Taavola},
  {Taboada}, {Tamburro}, {Ter-Antonyan}, {Te{\v{s}}i{\'c}}, {Tilav}, {Toale},
  {Toscano}, {Usner}, {van der Drift}, {van Eijndhoven}, {Van Overloop}, {van
  Santen}, {Vehring}, {Voge}, {Vraeghe}, {Walck}, {Waldenmaier}, {Wallraff},
  {Wasserman}, {Weaver}, {Wellons}, {Wendt}, {Westerhoff}, {Whitehorn},
  {Wiebe}, {Wiebusch}, {Williams}, {Wissing}, {Wolf}, {Wood}, {Woschnagg},
  {Xu}, {Xu}, {Xu}, {Yanez}, {Yodh}, {Yoshida}, {Zarzhitsky}, {Ziemann},
  {Zierke}, {Zilles}, \& {Zoll}}]{IceCube-2013PRL}
{Aartsen}, M.~G., {Abbasi}, R., {Abdou}, Y., {et~al.} 2013, \prl, 111, 021103,
  \dodoi{10.1103/PhysRevLett.111.021103}

\bibitem[{Aartsen {et~al.}(2019)}]{IceCube7yrCascades}
Aartsen, M.~G., {et~al.} 2019, Astrophys. J., 886, 12,
  \dodoi{10.3847/1538-4357/ab4ae2}

\bibitem[{{Abbasi} {et~al.}(2021){Abbasi}, {Ackermann}, {Adams}, {Aguilar},
  {Ahlers}, {Ahrens}, {Alispach}, {Alves}, {Amin}, {Andeen}, {Anderson},
  {Ansseau}, {Anton}, {Arg{\"u}elles}, {Axani}, {Bai}, {Balagopal V.},
  {Barbano}, {Barwick}, {Bastian}, {Basu}, {Baum}, {Baur}, {Bay}, {Beatty},
  {Becker}, {Becker Tjus}, {Bellenghi}, {BenZvi}, {Berley}, {Bernardini},
  {Besson}, {Binder}, {Bindig}, {Blaufuss}, {Blot}, {B{\"o}ser}, {Botner},
  {B{\"o}ttcher}, {Bourbeau}, {Bourbeau}, {Bradascio}, {Braun}, {Bron},
  {Brostean-Kaiser}, {Burgman}, {Busse}, {Campana}, {Chen}, {Chirkin}, {Choi},
  {Clark}, {Clark}, {Classen}, {Coleman}, {Collin}, {Conrad}, {Coppin},
  {Correa}, {Cowen}, {Cross}, {Dave}, {De Clercq}, {DeLaunay}, {Dembinski},
  {Deoskar}, {De Ridder}, {Desai}, {Desiati}, {de Vries}, {de Wasseige}, {de
  With}, {DeYoung}, {Dharani}, {Diaz}, {D{\'\i}az-V{\'e}lez}, {Dujmovic},
  {Dunkman}, {DuVernois}, {Dvorak}, {Ehrhardt}, {Eller}, {Engel}, {Evans},
  {Evenson}, {Fahey}, {Fazely}, {Fiedlschuster}, {Fienberg}, {Filimonov},
  {Finley}, {Fischer}, {Fox}, {Franckowiak}, {Friedman}, {Fritz}, {F{\"u}rst},
  {Gaisser}, {Gallagher}, {Ganster}, {Garrappa}, {Gerhardt}, {Ghadimi},
  {Glauch}, {Gl{\"u}senkamp}, {Goldschmidt}, {Gonzalez}, {Goswami}, {Grant},
  {Gr{\'e}goire}, {Griffith}, {Griswold}, {G{\"u}nd{\"u}z}, {Haack},
  {Hallgren}, {Halliday}, {Halve}, {Halzen}, {Ha Minh}, {Hanson}, {Hardin},
  {Haungs}, {Hauser}, {Hebecker}, {Helbing}, {Henningsen}, {Hickford},
  {Hignight}, {Hill}, {Hill}, {Hoffman}, {Hoffmann}, {Hoinka},
  {Hokanson-Fasig}, {Hoshina}, {Huang}, {Huber}, {Huber}, {Hultqvist},
  {H{\"u}nnefeld}, {Hussain}, {In}, {Iovine}, {Ishihara}, {Jansson},
  {Japaridze}, {Jeong}, {Jones}, {Joppe}, {Kang}, {Kang}, {Kang}, {Kappes},
  {Kappesser}, {Karg}, {Karl}, {Karle}, {Katori}, {Katz}, {Kauer},
  {Kellermann}, {Kelley}, {Kheirandish}, {Kim}, {Kin}, {Kintscher}, {Kiryluk},
  {Klein}, {Koirala}, {Kolanoski}, {K{\"o}pke}, {Kopper}, {Kopper}, {Koskinen},
  {Koundal}, {Kovacevich}, {Kowalski}, {Krings}, {Kr{\"u}ckl}, {Kulacz},
  {Kurahashi}, {Kyriacou}, {Lagunas Gualda}, {Lanfranchi}, {Larson}, {Lauber},
  {Lazar}, {Leonard}, {Leszczy{\'n}ska}, {Li}, {Liu}, {Lohfink}, {Lozano
  Mariscal}, {Lu}, {Lucarelli}, {Ludwig}, {Luszczak}, {Lyu}, {Ma}, {Madsen},
  {Mahn}, {Makino}, {Mallik}, {Mancina}, {Mandalia}, {Mari{\c{s}}}, {Maruyama},
  {Mase}, {McNally}, {Meagher}, {Medina}, {Meier}, {Meighen-Berger}, {Merz},
  {Micallef}, {Mockler}, {Moment{\'e}}, {Montaruli}, {Moore}, {Morse},
  {Moulai}, {Naab}, {Nagai}, {Naumann}, {Necker}, {Neer},
  {Nguy{\'a}{\guillemotright} n}, {Niederhausen}, {Nisa}, {Nowicki}, {Nygren},
  {Obertacke Pollmann}, {Oehler}, {Olivas}, {O'Sullivan}, {Pandya}, {Pankova},
  {Park}, {Parker}, {Paudel}, {Peiffer}, {P{\'e}rez de los Heros}, {Philippen},
  {Pieloth}, {Pieper}, {Pizzuto}, {Plum}, {Popovych}, {Porcelli}, {Prado
  Rodriguez}, {Price}, {Przybylski}, {Raab}, {Raissi}, {Rameez}, {Rawlins},
  {Rea}, {Rehman}, {Reimann}, {Renschler}, {Renzi}, {Resconi}, {Reusch},
  {Rhode}, {Richman}, {Riedel}, {Robertson}, {Roellinghoff}, {Rongen}, {Rott},
  {Ruhe}, {Ryckbosch}, {Rysewyk Cantu}, {Safa}, {Sanchez Herrera}, {Sandrock},
  {Sandroos}, {Santander}, {Sarkar}, {Sarkar}, {Satalecka}, {Scharf},
  {Schaufel}, {Schieler}, {Schlunder}, {Schmidt}, {Schneider}, {Schneider},
  {Schr{\"o}der}, {Schumacher}, {Sclafani}, {Seckel}, {Seunarine}, {Shefali},
  {Silva}, {Smithers}, {Snihur}, {Soedingrekso}, {Soldin}, {Spiczak},
  {Spiering}, {Stachurska}, {Stamatikos}, {Stanev}, {Stein}, {Stettner},
  {Steuer}, {Stezelberger}, {Stokstad}, {Strotjohann}, {Stuttard}, {Sullivan},
  {Taboada}, {Tenholt}, {Ter-Antonyan}, {Tilav}, {Tischbein}, {Tollefson},
  {Tomankova}, {T{\"o}nnis}, {Toscano}, {Tosi}, {Trettin}, {Tselengidou},
  {Tung}, {Turcati}, {Turcotte}, {Turley}, {Twagirayezu}, {Ty}, {Unger},
  {Unland Elorrieta}, {Vandenbroucke}, {van Eijk}, {van Eijndhoven},
  {Vannerom}, {van Santen}, {Verpoest}, {Vraeghe}, {Walck}, {Wallace},
  {Wandkowsky}, {Watson}, {Weaver}, {Weindl}, {Weiss}, {Weldert}, {Wendt},
  {Werthebach}, {Weyrauch}, {Whelan}, {Whitehorn}, {Wiebe}, {Wiebusch},
  {Williams}, {Wolf}, {Wood}, {Woschnagg}, {Wrede}, {Wulff}, {Xu}, {Xu},
  {Yanez}, {Yoshida}, {Yuan}, {Zhang}, \& {IceCube
  Collaboration}}]{IceCube-HESE2020}
{Abbasi}, R., {Ackermann}, M., {Adams}, J., {et~al.} 2021, \prd, 104, 022002,
  \dodoi{10.1103/PhysRevD.104.022002}

\bibitem[{{Abdalla} {et~al.}(2021){Abdalla}, {Aharonian}, {Ait Benkhali},
  {Ang{\"u}ner}, {Arcaro}, {Armand}, {Armstrong}, {Ashkar}, {Backes},
  {Baghmanyan}, {Barbosa Martins}, {Barnacka}, {Barnard}, {Becherini}, {Berge},
  {Bernl{\"o}hr}, {Bi}, {B{\"o}ttcher}, {Boisson}, {Bolmont}, {de Bony de
  Lavergne}, {Breuhaus}, {Brun}, {Brun}, {Bryan}, {B{\"u}chele}, {Bulik},
  {Bylund}, {Caroff}, {Carosi}, {Casanova}, {Chand}, {Chandra}, {Chen},
  {Cotter}, {Cury{\l}o}, {Damascene Mbarubucyeye}, {Davids}, {Davies}, {Deil},
  {Devin}, {Dirson}, {Djannati-Atai}, {Dmytriiev}, {Donath}, {Doroshenko},
  {Dreyer}, {Duffy}, {Dyks}, {Egberts}, {Eichhorn}, {Einecke}, {Emery},
  {Ernenwein}, {Feijen}, {Fegan}, {Fiasson}, {Fichet de Clairfontaine},
  {Fontaine}, {Funk}, {F{\"u}{\ss}ling}, {Gabici}, {Gallant}, {Giavitto},
  {Giunti}, {Glawion}, {Glicenstein}, {Grondin}, {Hahn}, {Haupt}, {Hermann},
  {Hinton}, {Hofmann}, {Hoischen}, {Holch}, {Holler}, {H{\"o}rbe}, {Horns},
  {Huber}, {Jamrozy}, {Jankowsky}, {Jankowsky}, {Jardin-Blicq}, {Joshi},
  {Jung-Richardt}, {Kasai}, {Kastendieck}, {Katarzynski}, {Katz}, {Khangulyan},
  {Kh{\'e}lifi}, {Klepser}, {Klu{\'z}niak}, {Komin}, {Konno}, {Kosack},
  {Kostunin}, {Kreter}, {Lamanna}, {Lemi{\`e}re}, {Lemoine-Goumard}, {Lenain},
  {Leuschner}, {Levy}, {Lohse}, {Lypova}, {Mackey}, {Majumdar}, {Malyshev},
  {Malyshev}, {Marandon}, {Marchegiani}, {Marcowith}, {Mares},
  {Mart{\'\i}-Devesa}, {Marx}, {Maurin}, {Meintjes}, {Meyer}, {Mitchell},
  {Moderski}, {Mohrmann}, {Montanari}, {Moore}, {Morris}, {Moulin}, {Muller},
  {Murach}, {Nakashima}, {Nayerhoda}, {de Naurois}, {Ndiyavala}, {Niemiec},
  {Oakes}, {O'Brien}, {Odaka}, {Ohm}, {Olivera-Nieto}, {de Ona Wilhelmi},
  {Ostrowski}, {Panny}, {Panter}, {Parsons}, {Peron}, {Peyaud}, {Piel}, {Pita},
  {Poireau}, {Priyana Noel}, {Prokhorov}, {Prokoph}, {P{\"u}hlhofer}, {Punch},
  {Quirrenbach}, {Raab}, {Rauth}, {Reichherzer}, {Reimer}, {Reimer}, {Remy},
  {Renaud}, {Rieger}, {Rinchiuso}, {Romoli}, {Rowell}, {Rudak}, {Ruiz-Velasco},
  {Sahakian}, {Sailer}, {Salzmann}, {Sanchez}, {Santangelo}, {Sasaki},
  {Scalici}, {Sch{\"a}fer}, {Sch{\"u}ssler}, {Schutte}, {Schwanke},
  {Seglar-Arroyo}, {Senniappan}, {Seyffert}, {Shafi}, {Shapopi},
  {Shiningayamwe}, {Simoni}, {Sinha}, {Sol}, {Specovius}, {Spencer},
  {Spir-Jacob}, {Stawarz}, {Sun}, {Steenkamp}, {Stegmann}, {Steinmassl},
  {Steppa}, {Takahashi}, {Tavernier}, {Taylor}, {Terrier}, {Thiersen},
  {Tiziani}, {Tluczykont}, {Tomankova}, {Trichard}, {Tsirou}, {Tuffs},
  {Uchiyama}, {van der Walt}, {van Eldik}, {van Rensburg}, {van Soelen},
  {Vasileiadis}, {Veh}, {Venter}, {Vincent}, {Vink}, {V{\"o}lk}, {Wadiasingh},
  {Wagner}, {Watson}, {Werner}, {White}, {Wierzcholska}, {Wun Wong},
  {Yusafzai}, {Zacharias}, {Zanin}, {Zargaryan}, {Zdziarski}, {Zech}, {Zhu},
  {Zorn}, {Zouari}, {{\.Z}ywucka}, \& {Acero}}]{2021arXiv210606405A}
{Abdalla}, H., {Aharonian}, F., {Ait Benkhali}, F., {et~al.} 2021, arXiv
  e-prints, arXiv:2106.06405.
\newblock \doarXiv{2106.06405}

\bibitem[{{Abeysekara} {et~al.}(2018){Abeysekara}, {Benbow}, {Bird}, {Brill},
  {Brose}, {Buckley}, {Chromey}, {Daniel}, {Falcone}, {Finley}, {Fortson},
  {Furniss}, {Gent}, {Gillanders}, {Hanna}, {Hassan}, {Hervet}, {Holder},
  {Hughes}, {Humensky}, {Kaaret}, {Kar}, {Kertzman}, {Kieda}, {Krause},
  {Krennrich}, {Kumar}, {Lang}, {Lin}, {Maier}, {Moriarty}, {Mukherjee},
  {O'Brien}, {Ong}, {Otte}, {Park}, {Petrashyk}, {Pohl}, {Pueschel}, {Quinn},
  {Ragan}, {Richards}, {Roache}, {Sadeh}, {Santander}, {Schlenstedt},
  {Sembroski}, {Sushch}, {Tyler}, {Vassiliev}, {Wakely}, {Weinstein}, {Wells},
  {Wilcox}, {Wilhelm}, {Williams}, {Williamson}, {Zitzer}, {VERITAS
  Collaboration}, {Acciari}, {Ansoldi}, {Antonelli}, {Arbet Engels}, {Baack},
  {Babi{\'c}}, {Banerjee}, {Barres de Almeida}, {Barrio}, {Becerra
  Gonz{\'a}lez}, {Bednarek}, {Bernardini}, {Berti}, {Besenrieder},
  {Bhattacharyya}, {Bigongiari}, {Biland}, {Blanch}, {Bonnoli}, {Busetto},
  {Carosi}, {Ceribella}, {Cikota}, {Colak}, {Colin}, {Colombo}, {Contreras},
  {Cortina}, {Covino}, {D'Elia}, {Da Vela}, {Dazzi}, {De Angelis}, {De Lotto},
  {Delfino}, {Delgado}, {Di Pierro}, {Do Souto Espi{\~n}era}, {Dom{\'\i}nguez},
  {Dominis Prester}, {Dorner}, {Doro}, {Einecke}, {Elsaesser}, {Fallah
  Ramazani}, {Fattorini}, {Fern{\'a}ndez-Barral}, {Ferrara}, {Fidalgo},
  {Foffano}, {Fonseca}, {Font}, {Fruck}, {Galindo}, {Gallozzi}, {Garc{\'\i}a
  L{\'o}pez}, {Garczarczyk}, {Gasparyan}, {Gaug}, {Giammaria}, {Godinovi{\'c}},
  {Guberman}, {Hadasch}, {Hahn}, {Herrera}, {Hoang}, {Hrupec}, {Inoue},
  {Ishio}, {Iwamura}, {Kubo}, {Kushida}, {Kuve{\v{z}}di{\'c}}, {Lamastra},
  {Lelas}, {Leone}, {Lindfors}, {Lombardi}, {Longo}, {L{\'o}pez},
  {L{\'o}pez-Oramas}, {Machado de Oliveira Fraga}, {Maggio}, {Majumdar},
  {Makariev}, {Mallamaci}, {Maneva}, {Manganaro}, {Mannheim}, {Maraschi},
  {Mariotti}, {Mart{\'\i}nez}, {Masuda}, {Mazin}, {Minev}, {Miranda},
  {Mirzoyan}, {Molina}, {Moralejo}, {Moreno}, {Moretti}, {Munar-Adrover},
  {Neustroev}, {Niedzwiecki}, {Nievas Rosillo}, {Nigro}, {Nilsson}, {Ninci},
  {Nishijima}, {Noda}, {Nogu{\'e}s}, {N{\"o}the}, {Paiano}, {Palacio},
  {Paneque}, {Paoletti}, {Paredes}, {Pedaletti}, {Pe{\~n}il}, {Peresano},
  {Persic}, {Prada Moroni}, {Prandini}, {Puljak}, {Garcia}, {Rhode},
  {Rib{\'o}}, {Rico}, {Righi}, {Rugliancich}, {Saha}, {Sahakyan}, {Saito},
  {Satalecka}, {Schweizer}, {Sitarek}, {{\v{S}}nidari{\'c}}, {Sobczynska},
  {Somero}, {Stamerra}, {Strzys}, {Suri{\'c}}, {Tavecchio}, {Temnikov},
  {Terzi{\'c}}, {Teshima}, {Torres-Alb{\`a}}, {Tsujimoto}, {van Scherpenberg},
  {Vanzo}, {Vazquez Acosta}, {Vovk}, {Will}, {Zari{\'c}}, \& {MAGIC
  Collaboration}}]{VERITAS_MAGIC}
{Abeysekara}, A.~U., {Benbow}, W., {Bird}, R., {et~al.} 2018, \apjl, 867, L19,
  \dodoi{10.3847/2041-8213/aae70e}

\bibitem[{{Abeysekara} {et~al.}(2021){Abeysekara}, {Albert}, {Alfaro},
  {Alvarez}, {Camacho}, {Arteaga-Vel{\'a}zquez}, {Arunbabu}, {Rojas},
  {Solares}, {Baghmanyan}, {Belmont-Moreno}, {BenZvi}, {Blandford}, {Brisbois},
  {Caballero-Mora}, {Capistr{\'a}n}, {Carrami{\~n}ana}, {Casanova}, {Cotti},
  {Le{\'o}n}, {De la Fuente}, {Hernandez}, {Dingus}, {DuVernois}, {Durocher},
  {D{\'\i}az-V{\'e}lez}, {Ellsworth}, {Engel}, {Espinoza}, {Fan}, {Fang},
  {Fleischhack}, {Fraija}, {Galv{\'a}n-G{\'a}mez}, {Garcia},
  {Garc{\'\i}a-Gonz{\'a}lez}, {Garfias}, {Giacinti}, {Gonz{\'a}lez}, {Goodman},
  {Harding}, {Hernandez}, {Hinton}, {Hona}, {Huang}, {Hueyotl-Zahuantitla},
  {H{\"u}ntemeyer}, {Iriarte}, {Jardin-Blicq}, {Joshi}, {Kieda}, {Lara}, {Lee},
  {Vargas}, {Linnemann}, {Longinotti}, {Luis-Raya}, {Lundeen}, {Malone},
  {Martinez}, {Martinez-Castellanos}, {Mart{\'\i}nez-Castro}, {Matthews},
  {Miranda-Romagnoli}, {Morales-Soto}, {Moreno}, {Mostaf{\'a}}, {Nayerhoda},
  {Nellen}, {Newbold}, {Nisa}, {Noriega-Papaqui}, {Olivera-Nieto}, {Omodei},
  {Peisker}, {P{\'e}rez Araujo}, {P{\'e}rez-P{\'e}rez}, {Ren}, {Rho},
  {Rosa-Gonz{\'a}lez}, {Ruiz-Velasco}, {Salazar}, {Greus}, {Sandoval},
  {Schneider}, {Schoorlemmer}, {Serna}, {Smith}, {Springer}, {Surajbali},
  {Tollefson}, {Torres}, {Torres-Escobedo}, {Ure{\~n}a-Mena}, {Weisgarber},
  {Werner}, {Willox}, {Zepeda}, {Zhou}, {De Le{\'o}n}, \&
  {{\'A}lvarez}}]{2021NatAs...5..465A}
{Abeysekara}, A.~U., {Albert}, A., {Alfaro}, R., {et~al.} 2021, Nature
  Astronomy, 5, 465, \dodoi{10.1038/s41550-021-01318-y}

\bibitem[{{Aharonian} {et~al.}(2019){Aharonian}, {Yang}, \& {de O{\~n}a
  Wilhelmi}}]{2019NatAs...3..561A}
{Aharonian}, F., {Yang}, R., \& {de O{\~n}a Wilhelmi}, E. 2019, Nature
  Astronomy, 3, 561, \dodoi{10.1038/s41550-019-0724-0}

\bibitem[{{Aharonian} {et~al.}(2006){Aharonian}, {Akhperjanian}, {Bazer-Bachi},
  {Beilicke}, {Benbow}, {Berge}, {Bernl{\"o}hr}, {Boisson}, {Bolz}, {Borrel},
  {Braun}, {Brown}, {B{\"u}hler}, {B{\"u}sching}, {Carrigan}, {Chadwick},
  {Chounet}, {Cornils}, {Costamante}, {Degrange}, {Dickinson},
  {Djannati-Ata{\"\i}}, {O'C. Drury}, {Dubus}, {Egberts}, {Emmanoulopoulos},
  {Espigat}, {Feinstein}, {Ferrero}, {Fiasson}, {Fontaine}, {Funk}, {Funk},
  {F{\"u}{\ss}ling}, {Gallant}, {Giebels}, {Glicenstein}, {Goret},
  {Hadjichristidis}, {Hauser}, {Hauser}, {Heinzelmann}, {Henri}, {Hermann},
  {Hinton}, {Hoffmann}, {Hofmann}, {Holleran}, {Horns}, {Jacholkowska}, {de
  Jager}, {Kendziorra}, {Kh{\'e}lifi}, {Komin}, {Konopelko}, {Kosack},
  {Latham}, {Le Gallou}, {Lemi{\`e}re}, {Lemoine-Goumard}, {Lohse}, {Martin},
  {Martineau-Huynh}, {Marcowith}, {Masterson}, {Maurin}, {McComb}, {Moulin},
  {de Naurois}, {Nedbal}, {Nolan}, {Noutsos}, {Orford}, {Osborne}, {Ouchrif},
  {Panter}, {Pelletier}, {Pita}, {P{\"u}hlhofer}, {Punch}, {Raubenheimer},
  {Raue}, {Rayner}, {Reimer}, {Reimer}, {Ripken}, {Rob}, {Rolland }, {Rowell},
  {Sahakian}, {Santangelo}, {Saug{\'e}}, {Schlenker}, {Schlickeiser},
  {Schr{\"o}der}, {Schwanke}, {Schwarzburg}, {Shalchi}, {Sol}, {Spangler},
  {Spanier}, {Steenkamp}, {Stegmann}, {Superina}, {Tavernet}, {Terrier},
  {Tluczykont}, {van Eldik}, {Vasileiadis}, {Venter}, {Vincent}, {V{\"o}lk},
  {Wagner}, \& {Ward}}]{Aharonian06}
{Aharonian}, F., {Akhperjanian}, A.~G., {Bazer-Bachi}, A.~R., {et~al.} 2006,
  \aap, 460, 743, \dodoi{10.1051/0004-6361:20065940}

\bibitem[{Albert {et~al.}(2018)}]{IceCubeANTARES-Gal}
Albert, A., {et~al.} 2018, \apjl, 868, L20, \dodoi{10.3847/2041-8213/aaeecf}

\bibitem[{{Amato}(2020)}]{2020arXiv200104442A}
{Amato}, E. 2020, arXiv e-prints, arXiv:2001.04442.
\newblock \doarXiv{2001.04442}

\bibitem[{{Amenomori} {et~al.}(2021){Amenomori}, {Bao}, {Bi}, {Chen}, {Chen},
  {Chen}, {Chen}, {Chen}, {Cirennima}, {Danzengluobu}, {Fang}, {Fang}, {Feng},
  {Feng}, {Feng}, {Gao}, {Gomi}, {Gou}, {Guo}, {Guo}, {He}, {He}, {Hibino},
  {Hotta}, {Hu}, {Hu}, {Huang}, {Jia}, {Jiang}, {Jiang}, {Jin}, {Kasahara},
  {Katayose}, {Kato}, {Kato}, {Kawata}, {Kozai}, {Kurashige}, {Labaciren},
  {Li}, {Li}, {Li}, {Li}, {Lin}, {Liu}, {Liu}, {Liu}, {Liu}, {Liu}, {Liu},
  {Liu}, {Lou}, {Lu}, {Meng}, {Munakata}, {Nakada}, {Nakamura}, {Nakazawa},
  {Nanjo}, {Ning}, {Nishizawa}, {Ohnishi}, {Ohura}, {Okukawa}, {Ozawa}, {Qian},
  {Qian}, {Qian}, {Qu}, {Saito}, {Sakata}, {Sako}, {Sako}, {Shao}, {Shibata},
  {Shiomi}, {Sugimoto}, {Takano}, {Takita}, {Tan}, {Tateyama}, {Torii},
  {Tsuchiya}, {Udo}, {Wang}, {Wang}, {Wangdui}, {Wu}, {Xu}, {Xue}, {Yamamoto},
  {Yang}, {Yao}, {Yin}, {Yokoe}, {Yu}, {Yuan}, {Zhai}, {Zhang}, {Zhang},
  {Zhang}, {Zhang}, {Zhang}, {Zhang}, {Zhang}, {Zhang}, {Zhao}, {Zhaxisangzhu},
  \& {Tibet AS<SUB>{\ensuremath{\gamma}}</SUB>
  Collaboration}}]{2021PhRvL.127c1102A}
{Amenomori}, M., {Bao}, Y.~W., {Bi}, X.~J., {et~al.} 2021, \prl, 127, 031102,
  \dodoi{10.1103/PhysRevLett.127.031102}

\bibitem[{{Arons}(2012)}]{2012SSRv..173..341A}
{Arons}, J. 2012, \ssr, 173, 341, \dodoi{10.1007/s11214-012-9885-1}

\bibitem[{Bednarek(2005)}]{prediction-nu-Bednarek2005}
Bednarek, W. 2005, Astrophys. J., 631, 466, \dodoi{10.1086/432411}

\bibitem[{{Bednarek}(2006)}]{2006MNRAS.368..579B}
{Bednarek}, W. 2006, \mnras, 368, 579, \dodoi{10.1111/j.1365-2966.2006.10121.x}

\bibitem[{{Bosch-Ramon}(2021)}]{bosch-ramon21}
{Bosch-Ramon}, V. 2021, \aap, 645, A86, \dodoi{10.1051/0004-6361/202039666}

\bibitem[{{Bykov} {et~al.}(2012){Bykov}, {Gehrels}, {Krawczynski}, {Lemoine},
  {Pelletier}, \& {Pohl}}]{B12}
{Bykov}, A., {Gehrels}, N., {Krawczynski}, H., {et~al.} 2012, \ssr, 173, 309.
\newblock \doarXiv{1205.2208}

\bibitem[{{Bykov} {et~al.}(2017){Bykov}, {Amato}, {Petrov}, {Krassilchtchikov},
  \& {Levenfish}}]{BSPWN_2017}
{Bykov}, A.~M., {Amato}, E., {Petrov}, A.~E., {Krassilchtchikov}, A.~M., \&
  {Levenfish}, K.~P. 2017, \ssr, 207, 235, \dodoi{10.1007/s11214-017-0371-7}

\bibitem[{{Bykov} {et~al.}(2015){Bykov}, {Ellison}, {Gladilin}, \&
  {Osipov}}]{2015MNRAS.453..113B}
{Bykov}, A.~M., {Ellison}, D.~C., {Gladilin}, P.~E., \& {Osipov}, S.~M. 2015,
  \mnras, 453, 113, \dodoi{10.1093/mnras/stv1606}

\bibitem[{{Camilo} {et~al.}(2009){Camilo}, {Ray}, {Ransom}, {Burgay},
  {Johnson}, {Kerr}, {Gotthelf}, {Halpern}, {Reynolds}, {Romani}, {Demorest},
  {Johnston}, {van Straten}, {Saz Parkinson}, {Ziegler}, {Dormody}, {Thompson},
  {Smith}, {Harding}, {Abdo}, {Crawford}, {Freire}, {Keith}, {Kramer},
  {Roberts}, {Weltevrede}, \& {Wood}}]{2009ApJ...705....1C}
{Camilo}, F., {Ray}, P.~S., {Ransom}, S.~M., {et~al.} 2009, \apj, 705, 1,
  \dodoi{10.1088/0004-637X/705/1/1}

\bibitem[{{Cao} {et~al.}(2021){Cao}, {Aharonian}, {An}, {Axikegu}, {Bai},
  {Bao}, {Bastieri}, {Bi}, {Bi}, {Cai}, {Cai}, {Cao}, {Chang}, {Chang},
  {Chang}, {Chen}, {Chen}, {Chen}, {Chen}, {Chen}, {Chen}, {Chen}, {Chen},
  {Chen}, {Chen}, {Chen}, {Chen}, {Chen}, {Cheng}, {Cheng}, {Cui}, {Cui},
  {Cui}, {Dai}, {Dai}, {Dai}, {Danzengluobu}, {della Volpe}, {D'Ettorre
  Piazzoli}, {Dong}, {Fan}, {Fan}, {Fan}, {Fang}, {Fang}, {Feng}, {Feng},
  {Feng}, {Feng}, {Gao}, {Gao}, {Gao}, {Gao}, {Ge}, {Geng}, {Gong}, {Gou},
  {Gu}, {Guo}, {Guo}, {Guo}, {Guo}, {Han}, {He}, {He}, {He}, {He}, {He}, {He},
  {Heller}, {Hor}, {Hou}, {Hou}, {Hu}, {Hu}, {Hu}, {Hu}, {Huang}, {Huang},
  {Huang}, {Huang}, {Huang}, {Ji}, {Ji}, {Jia}, {Jiang}, {Jiang}, {Jin},
  {Kuleshov}, {Levochkin}, {Li}, {Li}, {Li}, {Li}, {Li}, {Li}, {Li}, {Li},
  {Li}, {Li}, {Li}, {Li}, {Li}, {Li}, {Li}, {Li}, {Li}, {Liang}, {Liang},
  {Lin}, {Liu}, {Liu}, {Liu}, {Liu}, {Liu}, {Liu}, {Liu}, {Liu}, {Liu}, {Liu},
  {Liu}, {Liu}, {Liu}, {Liu}, {Liu}, {Long}, {Lu}, {Lv}, {Ma}, {Ma}, {Ma},
  {Mao}, {Masood}, {Mitthumsiri}, {Montaruli}, {Nan}, {Pang},
  {Pattarakijwanich}, {Pei}, {Qi}, {Ruffolo}, {Rulev}, {S{\'a}iz}, {Shao},
  {Shchegolev}, {Sheng}, {Shi}, {Song}, {Stenkin}, {Stepanov}, {Sun}, {Sun},
  {Sun}, {Tam}, {Tang}, {Tian}, {Wang}, {Wang}, {Wang}, {Wang}, {Wang}, {Wang},
  {Wang}, {Wang}, {Wang}, {Wang}, {Wang}, {Wang}, {Wang}, {Wang}, {Wang},
  {Wang}, {Wang}, {Wang}, {Wang}, {Wang}, {Wang}, {Wei}, {Wei}, {Wei}, {Wen},
  {Wu}, {Wu}, {Wu}, {Wu}, {Wu}, {Xi}, {Xia}, {Xia}, {Xiang}, {Xiao}, {Xiao},
  {Xin}, {Xin}, {Xing}, {Xu}, {Xu}, {Xue}, {Yan}, {Yang}, {Yang}, {Yang},
  {Yang}, {Yang}, {Yang}, {Yang}, {Yao}, {Yao}, {Ye}, {Yin}, {Yin}, {You},
  {You}, {Yu}, {Yuan}, {Zeng}, {Zeng}, {Zeng}, {Zeng}, {Zha}, {Zhai}, {Zhang},
  {Zhang}, {Zhang}, {Zhang}, {Zhang}, {Zhang}, {Zhang}, {Zhang}, {Zhang},
  {Zhang}, {Zhang}, {Zhang}, {Zhang}, {Zhang}, {Zhang}, {Zhang}, {Zhang},
  {Zhang}, {Zhang}, {Zhao}, {Zhao}, {Zhao}, {Zhao}, {Zhao}, {Zheng}, {Zheng},
  {Zhou}, {Zhou}, {Zhou}, {Zhou}, {Zhou}, {Zhou}, {Zhu}, {Zhu}, {Zhu}, {Zhu},
  \& {Zuo}}]{LHAASONat21}
{Cao}, Z., {Aharonian}, F.~A., {An}, Q., {et~al.} 2021, \nat, 594, 33,
  \dodoi{10.1038/s41586-021-03498-z}

\bibitem[{{Cerutti} {et~al.}(2011){Cerutti}, {Dubus}, {Malzac}, {Szostek},
  {Belmont}, {Zdziarski}, \& {Henri}}]{2011A&A...529A.120C}
{Cerutti}, B., {Dubus}, G., {Malzac}, J., {et~al.} 2011, \aap, 529, A120,
  \dodoi{10.1051/0004-6361/201116581}

\bibitem[{{Chernyakova} {et~al.}(2020{\natexlab{a}}){Chernyakova}, {Malyshev},
  {Blay}, {van Soelen}, \& {Tsygankov}}]{2020MNRAS.495..365C}
{Chernyakova}, M., {Malyshev}, D., {Blay}, P., {van Soelen}, B., \&
  {Tsygankov}, S. 2020{\natexlab{a}}, \mnras, 495, 365,
  \dodoi{10.1093/mnras/staa1181}

\bibitem[{{Chernyakova} {et~al.}(2020{\natexlab{b}}){Chernyakova}, {Malyshev},
  {Mc Keague}, {van Soelen}, {Marais}, {Martin-Carrillo}, \&
  {Murphy}}]{2020MNRAS.497..648C}
{Chernyakova}, M., {Malyshev}, D., {Mc Keague}, S., {et~al.}
  2020{\natexlab{b}}, \mnras, 497, 648, \dodoi{10.1093/mnras/staa1876}

\bibitem[{{Corbel} {et~al.}(2012){Corbel}, {Dubus}, {Tomsick}, {Szostek},
  {Corbet}, {Miller-Jones}, {Richards}, {Pooley}, {Trushkin}, {Dubois}, {Hill},
  {Kerr}, {Max-Moerbeck}, {Readhead}, {Bodaghee}, {Tudose}, {Parent}, {Wilms},
  \& {Pottschmidt}}]{CorbelCygX3flare12}
{Corbel}, S., {Dubus}, G., {Tomsick}, J.~A., {et~al.} 2012, \mnras, 421, 2947,
  \dodoi{10.1111/j.1365-2966.2012.20517.x}

\bibitem[{{Cristofari}(2021)}]{2021Univ....7..324C}
{Cristofari}, P. 2021, Universe, 7, 324, \dodoi{10.3390/universe7090324}

\bibitem[{{Dermer} \& {Menon}(2009)}]{2009herb.book.....D}
{Dermer}, C.~D., \& {Menon}, G. 2009, {High Energy Radiation from Black Holes:
  Gamma Rays, Cosmic Rays, and Neutrinos} (Princeton U.\ Press)

\bibitem[{Distefano {et~al.}(2002)Distefano, Guetta, Waxman, \&
  Levinson}]{prediction-nu-Waxman2002}
Distefano, C., Guetta, D., Waxman, E., \& Levinson, A. 2002, Astrophys. J.,
  575, 378, \dodoi{10.1086/341144}

\bibitem[{{Dubus}(2013)}]{2013A&ARv..21...64D}
{Dubus}, G. 2013, \aapr, 21, 64, \dodoi{10.1007/s00159-013-0064-5}

\bibitem[{{Dzhappuev} {et~al.}(2021){Dzhappuev}, {Afashokov}, {Dzaparova},
  {Dzhatdoev}, {Gorbacheva}, {Karpikov}, {Khadzhiev}, {Klimenko}, {Kudzhaev},
  {Kurenya}, {Lidvansky}, {Mikhailova}, {Petkov}, {Podlesnyi}, {Romanenko},
  {Rubtsov}, {Troitsky}, {Unatlokov}, {Vaiman}, {Yanin}, {Zhezher},
  {Zhuravleva}, \& {Carpet-3 Group}}]{Carpet2}
{Dzhappuev}, D.~D., {Afashokov}, Y.~Z., {Dzaparova}, I.~M., {et~al.} 2021,
  \apjl, 916, L22, \dodoi{10.3847/2041-8213/ac14b2}

\bibitem[{{Falanga} {et~al.}(2021){Falanga}, {Bykov}, {Li}, {Krassilchtchikov},
  {Petrov}, \& {Bozzo}}]{falanga21}
{Falanga}, M., {Bykov}, A.~M., {Li}, Z., {et~al.} 2021, arXiv e-prints,
  arXiv:2104.07711, \dodoi{10.1051/0004-6361/202141102}

\bibitem[{{HESS Collaboration} {et~al.}(2016){HESS Collaboration},
  {Abramowski}, {Aharonian}, {Benkhali}, {Akhperjanian}, {Ang{\"u}ner},
  {Backes}, {Balzer}, {Becherini}, {Tjus}, {Berge}, {Bernhard}, {Bernl{\"o}hr},
  {Birsin}, {Blackwell}, {B{\"o}ttcher}, {Boisson}, {Bolmont}, {Bordas},
  {Bregeon}, {Brun}, {Brun}, {Bryan}, {Bulik}, {Carr}, {Casanova},
  {Chakraborty}, {Chalme-Calvet}, {Chaves}, {Chen}, {Chr{\'e}tien},
  {Colafrancesco}, {Cologna}, {Conrad}, {Couturier}, {Cui}, {Davids},
  {Degrange}, {Deil}, {Dewilt}, {Djannati-Ata{\"\i}}, {Domainko}, {Donath},
  {Drury}, {Dubus}, {Dutson}, {Dyks}, {Dyrda}, {Edwards}, {Egberts}, {Eger},
  {Ernenwein}, {Espigat}, {Farnier}, {Fegan}, {Feinstein}, {Fernandes},
  {Fernandez}, {Fiasson}, {Fontaine}, {F{\"o}rster}, {F{\"u}{\ss}ling},
  {Gabici}, {Gajdus}, {Gallant}, {Garrigoux}, {Giavitto}, {Giebels},
  {Glicenstein}, {Gottschall}, {Goyal}, {Grondin}, {Grudzi{\'n}ska}, {Hadasch},
  {H{\"a}ffner}, {Hahn}, {Hawkes}, {Heinzelmann}, {Henri}, {Hermann}, {Hervet},
  {Hillert}, {Hinton}, {Hofmann}, {Hofverberg}, {Hoischen}, {Holler}, {Horns},
  {Ivascenko}, {Jacholkowska}, {Jamrozy}, {Janiak}, {Jankowsky},
  {Jung-Richardt}, {Kastendieck}, {Katarzy{\'n}ski}, {Katz}, {Kerszberg},
  {Kh{\'e}lifi}, {Kieffer}, {Klepser}, {Klochkov}, {Klu{\'z}niak}, {Kolitzus},
  {Komin}, {Kosack}, {Krakau}, {Krayzel}, {Kr{\"u}ger}, {Laffon}, {Lamanna},
  {Lau}, {Lefaucheur}, {Lefranc}, {Lemi{\'e}re}, {Lemoine-Goumard}, {Lenain},
  {Lohse}, {Lopatin}, {Lu}, {Lui}, {Marandon}, {Marcowith}, {Mariaud}, {Marx},
  {Maurin}, {Maxted}, {Mayer}, {Meintjes}, {Menzler}, {Meyer}, {Mitchell},
  {Moderski}, {Mohamed}, {Mor{\r{a}}}, {Moulin}, {Murach}, {de Naurois},
  {Niemiec}, {Oakes}, {Odaka}, {{\"O}ttl}, {Ohm}, {Opitz}, {Ostrowski}, {Oya},
  {Panter}, {Parsons}, {Arribas}, {Pekeur}, {Pelletier}, {Petrucci}, {Peyaud},
  {Pita}, {Poon}, {Prokoph}, {P{\"u}hlhofer}, {Punch}, {Quirrenbach}, {Raab},
  {Reichardt}, {Reimer}, {Reimer}, {Renaud}, {de Los Reyes}, {Rieger},
  {Romoli}, {Rosier-Lees}, {Rowell}, {Rudak}, {Rulten}, {Sahakian}, {Salek},
  {Sanchez}, {Santangelo}, {Sasaki}, {Schlickeiser}, {Sch{\"u}ssler}, {Schulz},
  {Schwanke}, {Schwemmer}, {Seyffert}, {Simoni}, {Sol}, {Spanier}, {Spengler},
  {Spies}, {Stawarz}, {Steenkamp}, {Stegmann}, {Stinzing}, {Stycz}, {Sushch},
  {Tavernet}, {Tavernier}, {Taylor}, {Terrier}, {Tluczykont}, {Trichard},
  {Tuffs}, {Valerius}, {van der Walt}, {van Eldik}, {van Soelen},
  {Vasileiadis}, {Veh}, {Venter}, {Viana}, {Vincent}, {Vink}, {Voisin},
  {V{\"o}lk}, {Vuillaume}, {Wagner}, {Wagner}, {Wagner}, {Weidinger},
  {Weitzel}, {White}, {Wierzcholska}, {Willmann}, {W{\"o}rnlein}, {Wouters},
  {Yang}, {Zabalza}, {Zaborov}, {Zacharias}, {Zdziarski}, {Zech}, {Zefi}, \&
  {{\.Z}ywucka}}]{GCPeV2016Natur}
{HESS Collaboration}, {Abramowski}, A., {Aharonian}, F., {et~al.} 2016, \nat,
  531, 476, \dodoi{10.1038/nature17147}

\bibitem[{{Ho} {et~al.}(2017){Ho}, {Ng}, {Lyne}, {Stappers}, {Coe}, {Halpern},
  {Johnson}, \& {Steele}}]{2017MNRAS.464.1211H}
{Ho}, W. C.~G., {Ng}, C.~Y., {Lyne}, A.~G., {et~al.} 2017, \mnras, 464, 1211,
  \dodoi{10.1093/mnras/stw2420}

\bibitem[{{IceCube Collaboration}(2013)}]{IceCube-2013Science}
{IceCube Collaboration}. 2013, Science, 342, 1242856,
  \dodoi{10.1126/science.1242856}

\bibitem[{{Kafexhiu} {et~al.}(2014){Kafexhiu}, {Aharonian}, {Taylor}, \&
  {Vila}}]{2014PhRvD..90l3014K}
{Kafexhiu}, E., {Aharonian}, F., {Taylor}, A.~M., \& {Vila}, G.~S. 2014, \prd,
  90, 123014, \dodoi{10.1103/PhysRevD.90.123014}

\bibitem[{{Kelner} \& {Aharonian}(2008)}]{2008PhRvD..78c4013K}
{Kelner}, S.~R., \& {Aharonian}, F.~A. 2008, \prd, 78, 034013,
  \dodoi{10.1103/PhysRevD.78.034013}

\bibitem[{{Kelner} {et~al.}(2006){Kelner}, {Aharonian}, \&
  {Bugayov}}]{2006PhRvD..74c4018K}
{Kelner}, S.~R., {Aharonian}, F.~A., \& {Bugayov}, V.~V. 2006, \prd, 74,
  034018, \dodoi{10.1103/PhysRevD.74.034018}

\bibitem[{Kheirandish(2020)}]{Kheirandish-Gal}
Kheirandish, A. 2020, Astrophys. Space Sci., 365, 108,
  \dodoi{10.1007/s10509-020-03816-3}

\bibitem[{{Klement} {et~al.}(2017){Klement}, {Carciofi}, {Rivinius},
  {Matthews}, {Vieira}, {Ignace}, {Bjorkman}, {Mota}, {Faes}, {Bratcher},
  {Cur{\'e}}, \& {{\v{S}}tefl}}]{Be_disk17}
{Klement}, R., {Carciofi}, A.~C., {Rivinius}, T., {et~al.} 2017, \aap, 601,
  A74, \dodoi{10.1051/0004-6361/201629932}

\bibitem[{{Lagunas Gualda} {et~al.}(2020)}]{IceCube-GCN.28927}
{Lagunas Gualda}, C., {et~al.} 2020, GRB Coordinates Network, 28927, 1

\bibitem[{{Lemoine} \& {Waxman}(2009)}]{2009JCAP...11..009L}
{Lemoine}, M., \& {Waxman}, E. 2009, \jcap, 2009, 009,
  \dodoi{10.1088/1475-7516/2009/11/009}

\bibitem[{Levinson \& Waxman(2001)}]{prediction-nu-Waxman2001}
Levinson, A., \& Waxman, E. 2001, Phys. Rev. Lett., 87, 171101,
  \dodoi{10.1103/PhysRevLett.87.171101}

\bibitem[{{Liu} \& {Kheirandish}(2021)}]{IceCube-binaries}
{Liu}, Q., \& {Kheirandish}, A. 2021, arXiv e-prints, arXiv:2107.12383.
\newblock \doarXiv{2107.12383}

\bibitem[{{Lyne} {et~al.}(2015){Lyne}, {Stappers}, {Keith}, {Ray}, {Kerr},
  {Camilo}, \& {Johnson}}]{2015MNRAS.451..581L}
{Lyne}, A.~G., {Stappers}, B.~W., {Keith}, M.~J., {et~al.} 2015, \mnras, 451,
  581, \dodoi{10.1093/mnras/stv236}

\bibitem[{{Murase} \& {Fukugita}(2019)}]{2019PhRvD..99f3012M}
{Murase}, K., \& {Fukugita}, M. 2019, \prd, 99, 063012,
  \dodoi{10.1103/PhysRevD.99.063012}

\bibitem[{{Neronov} \& {Ribordy}(2009)}]{prediction-nu-Neronov2009}
{Neronov}, A., \& {Ribordy}, M. 2009, \prd, 79, 043013,
  \dodoi{10.1103/PhysRevD.79.043013}

\bibitem[{{Neronov} \& {Semikoz}(2016)}]{NeronovSemikoz2comp}
{Neronov}, A., \& {Semikoz}, D. 2016, \prd, 93, 123002,
  \dodoi{10.1103/PhysRevD.93.123002}

\bibitem[{{Nikishov}(1962)}]{Nikishov1962}
{Nikishov}, A. 1962, Sov.\ Phys.\ JETP, 14, 393

\bibitem[{Palladino \& Vissani(2016)}]{2comp-Vissani}
Palladino, A., \& Vissani, F. 2016, Astrophys. J., 826, 185,
  \dodoi{10.3847/0004-637X/826/2/185}

\bibitem[{{Pittard} {et~al.}(2021){Pittard}, {Romero}, \& {Vila}}]{Pittard21}
{Pittard}, J.~M., {Romero}, G.~E., \& {Vila}, G.~S. 2021, \mnras, 504, 4204,
  \dodoi{10.1093/mnras/stab1107}

\bibitem[{{Romero} {et~al.}(2017){Romero}, {Boettcher}, {Markoff}, \&
  {Tavecchio}}]{Romero17}
{Romero}, G.~E., {Boettcher}, M., {Markoff}, S., \& {Tavecchio}, F. 2017, \ssr,
  207, 5, \dodoi{10.1007/s11214-016-0328-2}

\bibitem[{Sahakyan {et~al.}(2014)Sahakyan, Piano, \&
  Tavani}]{prediction-nu-Sahakyan2013}
Sahakyan, N., Piano, G., \& Tavani, M. 2014, Astrophys. J., 780, 29,
  \dodoi{10.1088/0004-637X/780/1/29}

\bibitem[{{Shultz} {et~al.}(2019){Shultz}, {Wade}, {Rivinius}, {Alecian},
  {Neiner}, {Petit}, {Owocki}, {ud-Doula}, {Kochukhov}, {Bohlender},
  {Keszthelyi}, {MiMeS Collaboration}, \& {BinaMIcS
  Collaboration}}]{2019MNRAS.490..274S}
{Shultz}, M.~E., {Wade}, G.~A., {Rivinius}, T., {et~al.} 2019, \mnras, 490,
  274, \dodoi{10.1093/mnras/stz2551}

\bibitem[{{Sironi} {et~al.}(2015){Sironi}, {Keshet}, \&
  {Lemoine}}]{2015SSRv..191..519S}
{Sironi}, L., {Keshet}, U., \& {Lemoine}, M. 2015, \ssr, 191, 519,
  \dodoi{10.1007/s11214-015-0181-8}

\bibitem[{{Takata} {et~al.}(2017){Takata}, {Tam}, {Ng}, {Li}, {Kong}, {Hui}, \&
  {Cheng}}]{2017ApJ...836..241T}
{Takata}, J., {Tam}, P.~H.~T., {Ng}, C.~W., {et~al.} 2017, \apj, 836, 241,
  \dodoi{10.3847/1538-4357/aa5c80}

\bibitem[{{Tavani} \& {Arons}(1997)}]{1997ApJ...477..439T}
{Tavani}, M., \& {Arons}, J. 1997, \apj, 477, 439, \dodoi{10.1086/303676}

\bibitem[{{The LHAASO collaboration}(2021)}]{2021arXiv210609865T}
{The LHAASO collaboration}. 2021, arXiv e-prints, arXiv:2106.09865.
\newblock \doarXiv{2106.09865}

\bibitem[{{Tibet AS{\ensuremath{\gamma}} Collaboration} {et~al.}(2021){Tibet
  AS{\ensuremath{\gamma}} Collaboration}, {Amenomori}, {Bao}, {Bi}, {Chen},
  {Chen}, {Chen}, {Chen}, {Chen}, {Cirennima}, {Danzengluobu}, {Fang}, {Fang},
  {Feng}, {Feng}, {Feng}, {Gao}, {Gou}, {Guo}, {Guo}, {He}, {He}, {Hibino},
  {Hotta}, {Hu}, {Hu}, {Huang}, {Jia}, {Jiang}, {Jin}, {Kasahara}, {Katayose},
  {Kato}, {Kato}, {Kawata}, {Kihara}, {Ko}, {Kozai}, {Labaciren}, {Li}, {Li},
  {Li}, {Lin}, {Liu}, {Liu}, {Liu}, {Liu}, {Liu}, {Lou}, {Lu}, {Meng},
  {Munakata}, {Nakada}, {Nakamura}, {Nanjo}, {Nishizawa}, {Ohnishi}, {Ohura},
  {Ozawa}, {Qian}, {Qu}, {Saito}, {Sakata}, {Sako}, {Shao}, {Shibata},
  {Shiomi}, {Sugimoto}, {Takano}, {Takita}, {Tan}, {Tateyama}, {Torii},
  {Tsuchiya}, {Udo}, {Wang}, {Wu}, {Xue}, {Yamamoto}, {Yang}, {Yokoe}, {Yuan},
  {Zhai}, {Zhang}, {Zhang}, {Zhang}, {Zhang}, {Zhang}, {Zhang}, {Zhang},
  {Zhao}, \& {Zhaxisangzhu}}]{2021NatAs...5..460T}
{Tibet AS{\ensuremath{\gamma}} Collaboration}, {Amenomori}, M., {Bao}, Y.~W.,
  {et~al.} 2021, Nature Astronomy, 5, 460, \dodoi{10.1038/s41550-020-01294-9}

\bibitem[{{Zdziarski} {et~al.}(2010){Zdziarski}, {Neronov}, \&
  {Chernyakova}}]{prediction-nu-Neronov2008}
{Zdziarski}, A.~A., {Neronov}, A., \& {Chernyakova}, M. 2010, \mnras, 403,
  1873, \dodoi{10.1111/j.1365-2966.2010.16263.x}

\end{thebibliography}
\bibliographystyle{aasjournal}

\end{document}